\PassOptionsToPackage{dvipsnames}{xcolor}
\PassOptionsToPackage{hyphens,obeyspaces,spaces}{url}

\documentclass[letterpaper,twocolumn,10pt]{article}

\usepackage{usenix2019_v3}

\usepackage{titlesec}
\titlespacing{\section}{0pt}{6pt plus 2pt minus 2pt}{4pt plus 2pt minus 2pt}
\titlespacing{\subsection}{0pt}{6pt plus 2pt minus 2pt}{4pt plus 2pt minus 2pt}
\titlespacing{\subsubsection}{0pt}{3pt plus 2pt minus 2pt}{2pt plus 1pt minus 1pt}
\titlespacing{\paragraph}{0pt}{\parskip}{-\parskip}

\usepackage[utf8]{inputenc}
\usepackage{array}
\newcolumntype{P}[1]{>{\centering\arraybackslash}p{#1}}
\newcolumntype{H}{>{\setbox0=\hbox\bgroup}c<{\egroup}@{}}

\usepackage{multirow}
\usepackage{makecell}
\usepackage{subcaption}
\usepackage{balance}
\usepackage{fancyhdr}
\usepackage{algorithm2e}
\usepackage{listings}
\usepackage{cprotect}
\usepackage{environ}

\usepackage[printwatermark]{xwatermark}
\newwatermark[allpages,color=black!100,angle=0,scale=1,xpos=0,ypos=-126]{\small 2021 30th USENIX Security Symposium \\ Accepted version. \url{https://www.usenix.org/conference/usenixsecurity21/presentation/jeitner}}

\usepackage{pifont}

\newcommand{\cmark}{\ding{51}} %
\newcommand{\xmark}{\ding{55}} %

\usepackage{tikz}

\newcommand*\circled[1]{\tikz[baseline=(char.base)]{
            \node[shape=circle,draw,inner sep=1pt,semithick] (char) {#1};}}

\newcommand{\stt}[1]{{\small\path{#1}}}
\newcommand{\name}{injection}
\newcommand{\new}[1]{#1}

\begin{document}

\date{}

\title{Injection Attacks Reloaded:\\Tunnelling Malicious Payloads over DNS}

\author{
{\rm Philipp\ Jeitner}\\
TU Darmstadt
\and
{\rm Haya Shulman}\\
Fraunhofer SIT
}

\maketitle

\begin{abstract}

The traditional design principle for Internet protocols indicates: ``Be strict when sending and tolerant when receiving'' [RFC1958], and DNS is no exception to this. The transparency of DNS in handling the DNS records, also standardised specifically for DNS [RFC3597], is one of the key features that made it such a popular platform facilitating a constantly increasing number of new applications. An application simply creates a new DNS record and can instantly start distributing it over DNS without requiring any changes to the DNS servers and platforms. Our Internet wide study confirms that more than 1.3M (96\% of tested) open DNS resolvers are standard compliant and treat DNS records transparently. 

In this work we show that this `transparency' introduces a severe vulnerability in the Internet: we demonstrate a new method to launch string injection attacks by encoding malicious payloads into DNS records. We show how to weaponise such DNS records to attack popular applications. For instance, we apply string injection to launch a new type of DNS cache poisoning attack, which we evaluated against a population of open resolvers and found 105K to be vulnerable. Such cache poisoning cannot be prevented with common setups of DNSSEC. Our attacks apply to internal as well as to public services, for instance, we reveal that all eduroam services are vulnerable to our injection attacks, allowing us to launch exploits ranging from unauthorised access to eduroam networks to resource starvation. Depending on the application, our attacks cause system crashes, data corruption and leakage, degradation of security, and can introduce remote code execution and arbitrary errors. 

In our evaluation of the attacks in the Internet we find that all the standard compliant open DNS resolvers we tested allow our injection attacks against applications and users on their networks.

\end{abstract}

\section{Introduction}

Domain Name System (DNS) is a key component of the Internet. Originally designed to translate domain names to IP addresses, DNS has developed into a complex infrastructure providing platform to a constantly increasing number of applications. The applications that are built over DNS range from Internet specific services, such as location of hosts using GPOS record [RFC1712] \cite{rfc1712} to security mechanisms, such as authentication with certificates using TLSA record [RFC6698] \cite{rfc6698}. The core design feature that allows DNS to support new applications without involving any changes to its infrastructure is the requirement that the handling of the DNS records is done transparently [RFC3597, RFC1035] \cite{rfc3597,rfc1035}. Namely, DNS should not attempt to interpret nor understand the records that it is serving. Thanks to this feature new DNS records can be easily added to the DNS infrastructure without requiring any modifications, and novel applications can instantly run over DNS using the newly added records.

In this work we show that the transparency-feature of DNS, while critical for fast and smooth deployment of new technologies, introduces a gaping hole in Internet security. %

{\bf Exploiting transparency to encode injections.} We exploit the transparency of the DNS lookups to encode injection strings into the payloads of DNS records. The attacker places the malicious records in the zonefile of its domain. When provided by the attacker's nameserver the records appear to contain legitimate mappings under the domain controlled by the attacker. However, when the record is processed by the receiving victim application, a misinterpretation occurs - resulting in the injection attack. Our attacks exploit two key factors caused by the transparency of DNS: (1) the DNS resolvers do not alter the received records hence the malicious encoding is preserved intact and (2) the receiving applications do not sanitise the received records. We devise injection payloads to attack popular applications.

{\bf Applications do not sanitise DNS records.} Classical injection attacks are well known and have been extensively studied: the attacker provides a malicious input through a web application to alter the structure of a command, hence subverting the logic of the application, e.g., \cite{halfond2006classification,grossman2007xss}.
Such injection attacks are easy to mitigate in practice: the input of the user is validated and invalid characters are filtered before reaching the application. Due to the long history of injection vulnerabilities in web applications and the awareness to the potential risks, most applications validate user input \cite{pietraszek2005defending}. 

We show that in contrast to user input, the inputs provided by the DNS resolvers are not validated. For instance, user credentials provided to LDAP via a web interface to authenticate the user and enable it to use services, are validated, while DNS values that are provided to LDAP to route the authentication request to an authentication server are not validated. We show how to construct malicious payloads to launch injection attacks, such as XSS and cache poisoning, against a variety of applications and services, including DNS caches, LDAP, eduroam. %

\begin{figure}[t]
    \centering
    \vspace{-10pt}
    \includegraphics[width=0.40\textwidth]{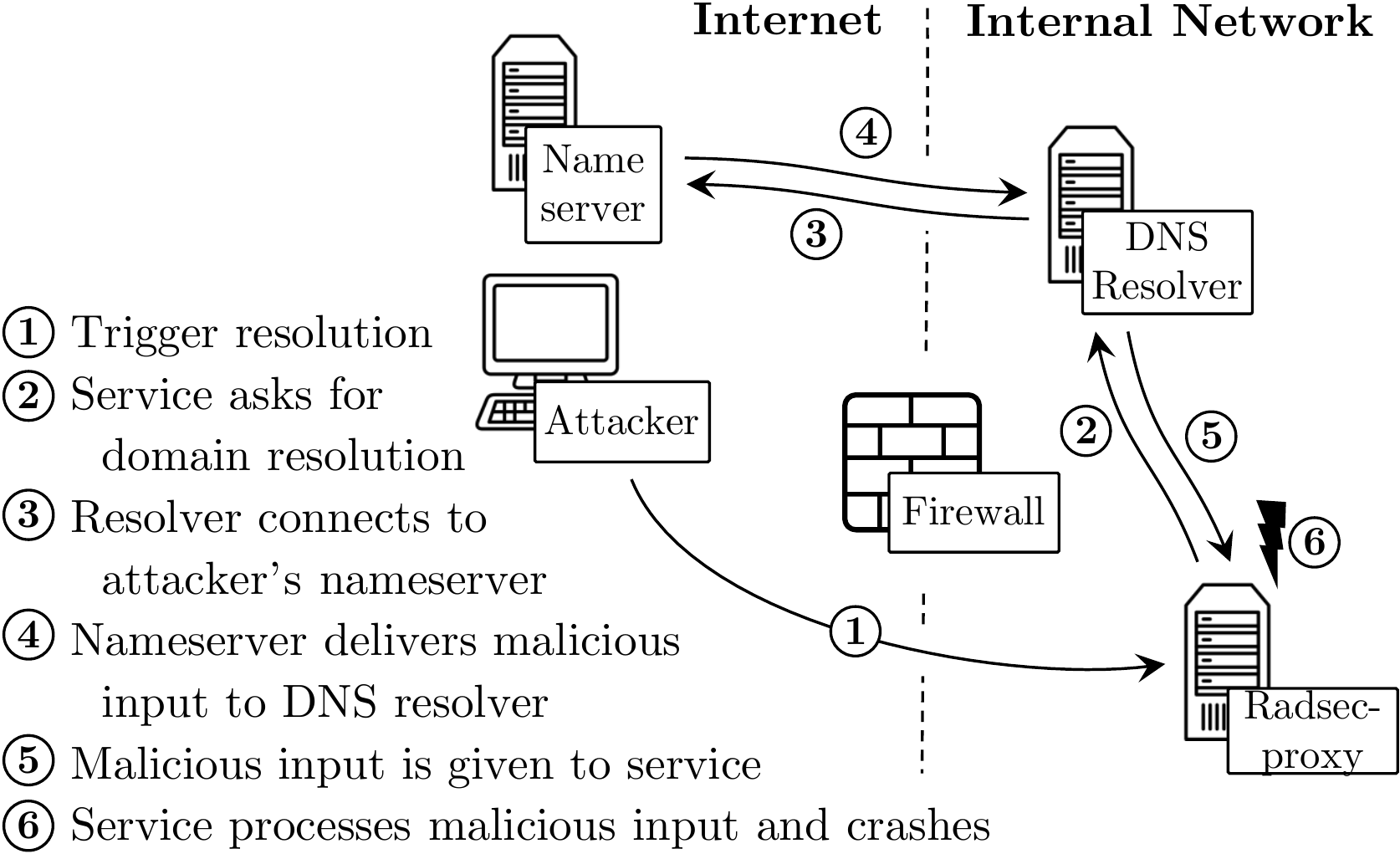}
    \caption{Attack and setup overview, with eduroam radsecproxy as example application.}
    \label{fig:attackoverview}
    \vspace{-14pt}
\end{figure}

\textbf{Attacker model.} The attacker causes the victim resolvers to issue queries for records that encode malicious payloads, e.g., by deploying an ad-network or by sending an Email from attacker's domain to the victim. The resolvers cache the records received in DNS responses and provide them to applications and users. We illustrate the attacker model and the setup with eduroam as example victim application, in Figure \ref{fig:attackoverview}. Using our ``weak'' attacker we demonstrate a range of attacks against popular applications and services that use DNS lookups, including DNS cache poisoning, applications' crashes, downgrade of security mechanisms, remote code execution vulnerabilities, XSS. %

\subsection*{Contributions} %
The core issue that we explore in this work is the balance between security and the requirement to enable easy deployment of new applications over DNS. Our contributions include:

$\bullet$ {\bf Analysis of components in resolution chain.} We analyse the interaction between the applications and the components in DNS resolution chain. We find that the processing applied by the DNS resolvers over DNS records is compliant with the requirement in [RFC3597,RFC1035] and preserves the structure of the malicious inputs encoded by the attackers - this property is key to our attacks. We validate this also in the Internet against 3M open DNS resolvers. Our measurement study reveals that more than 96\% of the open DNS resolvers do not modify the records that they receive from the nameservers, and serve them intact to the calling applications. %

$\bullet$ {\bf Study of DNS input validation.} We find that although DNS delivers untrusted data from potentially malicious Internet servers the applications trust the data returned by the DNS resolvers. Our study shows that the lack of input validation is systematic and prevalent and is not a bug mistakenly introduced by developers in some isolated cases -- this includes custom functions in applications as well as standardised function calls of IEEE POSIX, e.g.,  {\tt gethostbyname()}. %

$\bullet$ {\bf Implementation of injection attacks over DNS.} We demonstrate that the attackers can systematically and efficiently construct attack vectors and show how to integrate them into the zonefile of a malicious domain operated by the attacker. We then demonstrate \name\ attacks over DNS against popular applications using these malicious DNS records. %

$\bullet$ {\bf Injections into DNS caches.} We show how to encode payloads for injecting malicious records into DNS caches. When cached by the victim DNS resolver, a misinterpretation occurs mapping a resource of some victim domain to an attacker's IP address. In contrast to classical cache poisoning, \cite{cns:frag:dns,zheng2020poison,qian:ccs20}, which require a strong (on-path) attacker or assume specific network properties, such as side channels or fragmentation, our cache poisoning attacks do not make any requirements on attacker capabilities. We also do not need to spoof IP addresses in DNS responses - a requirement which is essential in prior cache poisoning attacks. %
We automated the evaluation of our poisoning attacks, which allowed us to launch them against a large set of 3M target DNS resolvers, performing successful poisoning against 105K resolvers. Implementation of previous cache poisoning attacks had to be manually tailored per each target - automating the attacks would result in a negligible success probability. Hence the previous attacks were carried out against at most a handful of targets, and the rest of the resolvers' population was merely checked for properties that make them potential targets. Furthermore, in contrast to previous poisoning attacks, ours cannot be prevented with common setups of DNSSEC \cite{rfc4033,rfc4034,rfc4035}.

$\bullet$ {\bf Evaluation of injection attacks over DNS.} We evaluate our \name\ attacks against popular applications (listed in Table \ref{tab:analyzed-apps}). Our analysis of the vulnerabilities in applications, where suitable, combines fuzzing, source code review and dynamic (black box) execution. %
We evaluate our \name\ attacks against a population of more than 3M open resolvers in the Internet. 
\new{We provide additional information on our evaluations at {\small\path{ https://xdi-attack.net}}.} %

\subsection*{Ethics and Disclosure} We have already taken preliminary steps to address these vulnerabilities by contacting the DNS software vendors as well the applications evaluated in this work. We experimentally evaluated the attacks reported in this work against servers that we set up as well as against open DNS resolvers in the Internet using domains that we control. This allowed us to validate the presence of the vulnerabilities without exploiting them against real victims and without causing damage to the networks nor services in the Internet. Our attacks similarly apply also to non-open DNS resolvers. %

Prior to performing the validation of the vulnerabilities in the wild we received an approval from our research institution. In the next steps we will be coordinating countermeasures with the DNS and applications vendors, as well as the IETF community.

\subsection*{Organisation} In Section \ref{sc:background} we analyse the interaction between components in DNS resolution chain. In Section \ref{sec:appvulnerabilities}, we demonstrate injection attacks against popular applications. In Section \ref{sec:resolverstudy} we evaluate our attacks against open resolvers in the Internet. We propose countermeasures in Section \ref{sc:mitigations}, review related work in Section \ref{sc:works} and conclude in Section \ref{sc:conclusions}.

\section{Analysis of DNS Resolution Chain}\label{sc:background} %

In this section we analyse the interaction between the components relevant to processing DNS records in responses. In our analysis we use popular DNS resolvers and stub resolver implementations built into operating systems and programming languages, and experimentally test how they handle control characters in domain names and if they modify any of the maliciously crafted payloads needed to conduct the application-specific exploits that we evaluate in this work. %

\begin{figure}
    \centering
    \includegraphics[width=0.46\textwidth]{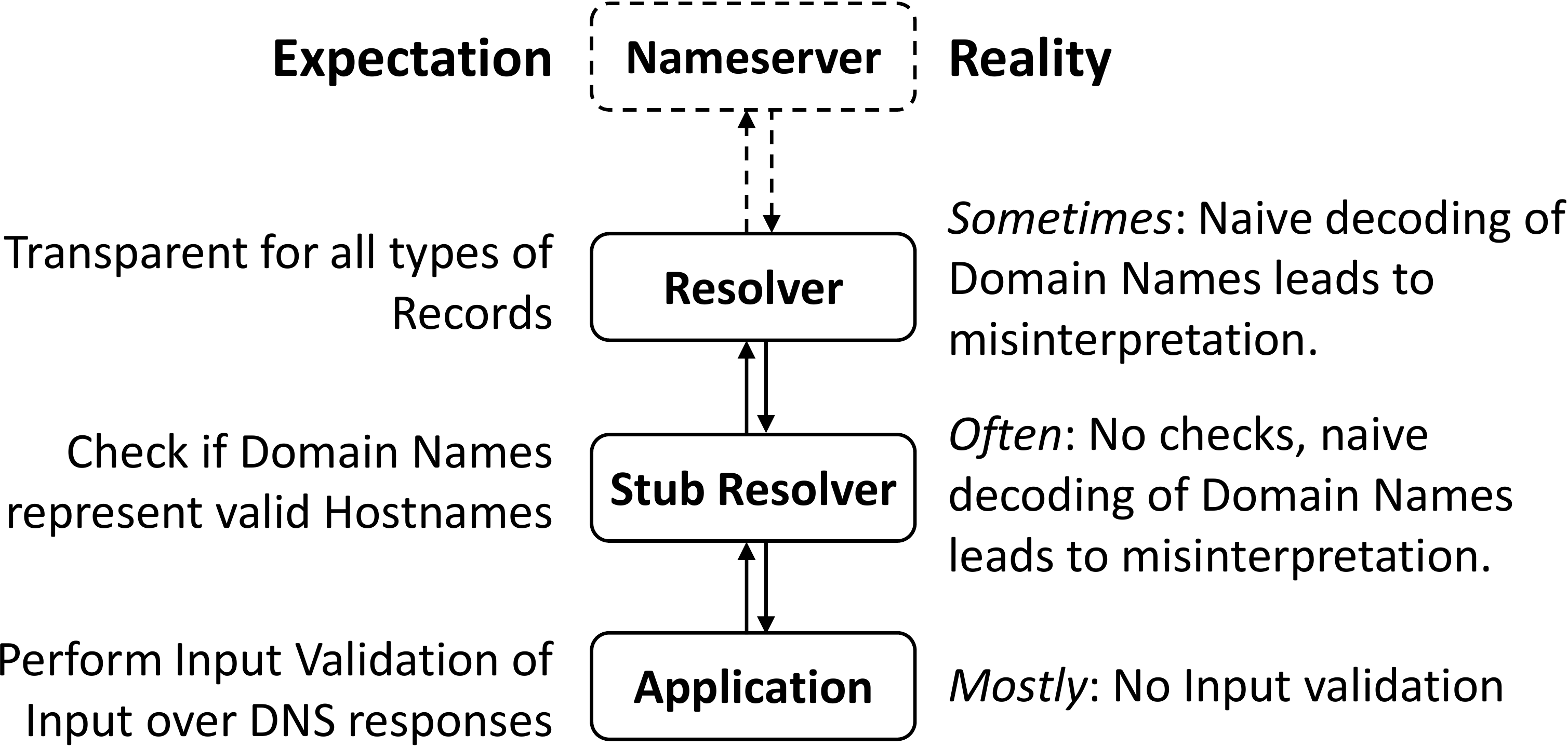}
    \caption{Expected vs. actual behaviour in DNS lookup.}
    \label{fig:expectation-reality-overview}
    \vspace{-7pt}
\end{figure}

\subsection{Components in DNS Lookup}
We consider 3 different types of software components which fulfil different roles during a DNS lookup: (recursive) DNS resolvers, stub DNS resolvers and applications. In our setup the DNS namerservers are controlled by the attacker and provide maliciously-encoded DNS responses. The victim resolvers serve these records to the stub resolvers in applications. We provide an illustration of these software components together with the expectations on their behaviour and their actual behaviour discovered in this work in  Figure~\ref{fig:expectation-reality-overview}.
DNS lookups by system stub resolvers are implemented in various ways in applications. For standard A, AAAA and PTR queries, system C libraries include POSIX-standardised \cite{POSIX.1-2008,rfc3493} functionality in form of the {\small\path{gethostbyname()}}, {\small\path{getaddrinfo()}}, {\small\path{gethostbyaddr()}} and {\small\path{getnameinfo()}} functions. When considering full fledged resolution functionality, which also supports other query types, like MX, SRV or TXT, there is no standardised API so applications need to rely on third party libraries for constructing the DNS packets or for parsing the DNS responses they receive from the network (e.g., recursive resolvers). DNS software implementations in recursive resolvers and forwarders typically implement their own packet decoding logic since they do not interface with the applications directly and do not need to decode parts of the DNS records at all.

\subsection{System Stub Resolvers}

Stub resolvers provide the interface between the applications and the DNS resolvers. Applications typically do not issue DNS requests to the recursive resolvers directly, but instead use a POSIX standardised API \cite{POSIX.1-2008,rfc3493} to supply the hostname they want to resolve, and the system's standard C library translates this into a DNS request and parses the response from the recursive DNS resolver. The applications can perform hostname-to-address (A, AAAA) and reverse (PTR) lookups via the {\small\path{gethostbyname()}}, {\small\path{gethostbyaddr()}}, {\small\path{getaddrinfo()}} and {\small\path{getnameinfo()}} calls.
This API is defined to return CNAME aliases and results of reverse lookups as null-terminated `host names' \cite[p.~320]{POSIX.1-2008} in its returned {\small\path{hostent}} structure, defined \cite{rfc952,rfc1123} to only contain Latin characters, digits and hyphens ({\small\path{ "A-Z","a-z","0-9","-"}}).

According to this definition the expected behaviour of a system stub resolver is to check that any domain name returned by a call to any of the POSIX-standardised resolver functions must be checked before returning it to the application. 

\new{\textbf{DNS Record Processing.}
We analyse the DNS record processing done by stub-resolvers in Section~\ref{sec:recorprocessing} in detail. We find that while one implementation (glibc) is fully conforming with our expectations, most stub-resolvers are not.} Stub resolvers that do not validate the value of the hostname expose the applications to attacks; we show these in Section \ref{sc:app:admintools}.

\subsection{Recursive Resolvers and Forwarders}
Recursive resolvers provide lookup services to system DNS resolvers: they locate nameservers and look up the requested records. %

DNS resolvers and forwarders process DNS records in their line-format, i.e., they store and handle domain names in their encoded form and do not try to parse the records that they do not need to understand, like TXT or MX. This makes them transparent for any binary values inside domains or other record data as is required by the DNS standard \cite{rfc1035,rfc4343,rfc6895}. For internal caching purposes, a DNS resolver may choose to decode a domain name for further processing or to store it inside a cache.

\textbf{DNS Record Processing.}
DNS resolvers are expected to transparently handle known and unknown DNS record types [RFC3597] to ensure forward-compatibility and compatibility with mechanisms such as DNSSEC, as any change in the encoded record would invalidate the DNSSEC signatures. Moreover, [RFC1034] states that 
software like DNS resolvers should not try to decode domain names into a string \new{as stub-resolvers would do}. Most resolvers we tested are [RFC1034] compliant: they handle any payload transparently or only change the case of letters to lowercase, which is allowed since domain names are defined to be case insensitive \cite{rfc4343}. One tested resolver software, MaraDNS Deadwood 3.2.14, was unable to handle inject\textsubscript{\textbackslash000}. %

We also identified an `[RFC1034] non-compliant' behaviour, which we describe next. We take as an example a systemd-resolved forwarding DNS resolver which when receiving a DNS record decodes and escapes all included domain names into zero-terminated strings like a system stub resolver would do. Then it caches the records as decoded strings. The resolver unescapes and re-encodes the records when sending them to a requesting client or application. %
This behaviour ensures that any misinterpretation during \new{decoding} (step 2 in Figure~\ref{fig:decode-flow}) will cause the domain name to be modified. This modification cannot be detected by downstream resolvers or applications because the misinterpreted record is re-encoded instead of just being passed as binary.  %

\subsection{Applications}
Applications are the source of DNS lookups and process the records in the response after it has traversed all the other components in the resolution chain. The records may contain unexpected characters which the application cannot process correctly. The impact of \name\ attacks on the applications depends on the use case of the DNS lookup. 
For example, applications which do service discovery or authentication lookups typically need to parse DNS packets by themselves, because system stub resolvers do not support queries for such record types. This can also be abused to cause misinterpretation of domain names or other data in DNS records.

\textbf{DNS Record Processing.}
When applications perform DNS lookups of types other than A, AAAA or PTR, they implement DNS lookup functionality by themselves as there is no standardised API for this. While the standard libraries of some programming languages like java, go or nodejs include functions for query types like MX, SRV and TXT, there is no recommended behaviour for such functions and they do not perform any validation of the data which is passed to the application. This also applies to DNS lookup implementations done in applications directly. Domain names are typically not validated nor escaped when decoded into a string.

Theoretically, when applications use the system resolver to do DNS lookups, they could implement validation (step 3 in Figure~\ref{fig:decode-flow}) by themselves, to ensure no malicious input is processed. However, applications would still not be able to detect decoding errors in step 2 in Figure~\ref{fig:decode-flow}. For example, since the application does not see the binary DNS data, it cannot determine if the example domain in Figure~\ref{fig:decode-flow} is \stt{"a\.b.<>.com."} or \stt{"a.b.<>.com."}.

\section{Injection Attacks Against Applications}
\label{sec:appvulnerabilities}
In this section we demonstrate \name\ attacks using malicious payloads tunnelled over DNS.
We first explain our study methodology (Section \ref{sec:appstudymethodology}) and then show attacks against selected popular applications (Sections \ref{sec:app-resolver} -- \ref{sec:app-admin}) taking DNS software as the first example application.  %

\subsection{Study Methodology}
\label{sec:appstudymethodology}

\subsubsection{Attack  \new{overview}}

The attack  \new{is} illustrated in Figure \ref{fig:attackoverview}. The target victim application, e.g., radsecproxy of eduroam, is behind a firewall on the victim network. The attack is initiated by causing the target service to issue a request via its DNS resolver to the attacker's domain, e.g., by trying to authenticate at eduroam's wireless access point (steps \circled{1}, \circled{2} and \circled{3} in Figure \ref{fig:attackoverview}). In the zonefile of its domain, the attacker encodes malicious payloads into the DNS records. The records are then provided in responses to the queries of the DNS resolvers (step \circled{4}), and are subsequently relayed to the requesting services, in the example in Figure \ref{fig:attackoverview}, to the radsecproxy server (step \circled{5}) which  processes the attacker's authentication request. The payload then causes the application to divert from a standard behaviour (step \circled{6}), e.g., causing it to allow unauthenticated network access. 

\subsubsection{Selecting Target Applications}
We evaluate \name\ attacks against popular services and applications. In this work we present attacks against some selected applications listed in Table~\ref{tab:analyzed-apps}. We select them based on the following considerations:

\textbf{DNS Use-Case.} We identify 4 different use-cases of DNS (address lookup, service discovery, reverse lookup and authentication). We select a few popular applications and services for each DNS use case. %

\textbf{Triggering query.} The attacker must be able to trigger a DNS lookup, e.g., via a script in a browser, via an Email to a target Email Server. We summarise methods for triggering query and setting the query domain in Table~\ref{tab:analyzed-apps}, column 'Trigger/Set query'. \new{We also prefer target applications which allow the attacker to trigger queries to attacker-selected domains, e.g., by sending an Email or triggering a query via javascript in browsers.}

\textbf{Attack surface.} To find meaningful attacks, we focus on applications where input from DNS is used for some interesting action, e.g., for implementing a cache, creating a URL, etc. \new{We do not analyse applications which only do standard address lookups (without caching), as such a scenario does not create a meaningful attack surface, even if no input validation is performed.} We list the applications, along with how the DNS inputs are used by those applications, in Table~\ref{tab:analyzed-apps}, column 'Input use'.

\new{\textbf{Usage of vulnerable resolvers.} For applications which use the system's libc resolver for DNS lookups, we prefer those which are often used on systems with vulnerable libc implementations. For example, OpenWRT was chosen because it uses a vulnerable libc implementation with uclibc.}

\subsubsection{Vulnerabilities Analysis}
After identifying a target application, we analyse its DNS usage and whether input from the DNS is validated, as follows: (1) source code review, (2) fuzzing and (3) by executing the application, feeding it with inputs and analysing the resulting behaviour and the outputs. We first test if an application does not validate DNS records received in input. For such applications we then check how the input is used by the application, and construct attack vectors accordingly, e.g., XSS injection. The results of this analysis are listed in Table~\ref{tab:analyzed-apps}, the found vulnerabilities in Table~\ref{tab:vulnerabilities}. %

\begin{table}[t]
    \footnotesize
    \centering
    \newcommand{\mc}[1]{\multicolumn{2}{c|}{#1}}
    \setlength{\tabcolsep}{1.3pt}
    \begin{tabular}{|lH|l|c|c|c|c|c|c|}
        \hline
        {\bf DNS Use-}            & Category              & {\bf Application}   & {\bf Trigger}   & {\bf Set}       & {\bf Uses}      & {\bf Vali-}     & {\bf Input}          & {\bf Attack} \\ \cline{4-5}
        {\bf Case}                &                       &               & \mc{Query}            & {\bf libc}      & {\bf dates}     & {\bf use}            & {\bf found} \\
        \hline
        Address             & Browser               & Chrome        & \mc{js,html}          & yes       & no        & cache          & no \\ %
        lookups             &                       & Firefox       & \mc{js,html}          & yes       & no        & cache          & no \\ %
        (A, CNAME)          &                       & Opera         & \mc{js,html}          & yes       & no        & cache          & no \\ %
                            &                       & Edge          & \mc{js,html}          & yes       & no        & cache          & no \\ %
                            & Cache                 & unscd         & \mc{client app}       & yes       & no        & cache          & no \\ %
                            &                       & java          & \mc{client app}       & both      & no        & cache          & no \\ \cline{4-5} %
                            & Tools                 & ping(win32)   & \xmark    & \xmark    & yes       & no        & display        & yes \\
        \hline
        discovery           & LDAP                  & openjdk       & login     & \xmark    & no        & no        & create URL     & yes   \\ %
        (MX, SRV,           &                       & ldapsearch    & login     & \xmark    & no        & no        & create URL     & no    \\ \cline{4-5} %
          NAPTR)            & Radius                & radsecproxy   & \mc{login}            & no        & no        & configure      & yes \\
        \hline
        Reverse             & Tools                 & ping(linux)   & \xmark    & \xmark    & yes       & no        & display        & yes \\ %
        lookups             &                       & trace(linux)  & \xmark    & \xmark    & yes       & no        & display        & yes \\ %
        (PTR)               & Web IF.               & OpenWRT       & \xmark    & ping      & yes       & no        & display        & yes \\ \cline{4-5} %
                            & SSH                   & openssh       & \mc{login}            & yes       & no        & display,log    & yes \\ %
        \hline
        Authentication      & SPF                   & policyd-spf   & \mc{SMTP}             & no        & no        & text protocol  & no \\ %
        (TXT, TLSA)         &                       & libspf2       & \mc{SMTP}             & no        & -         & parse          & yes \\ %
        \hline
        
        All                 & DNS                   & Resolvers     & \mc{client app}       & no        & some      & cache          & yes \\
        
        \hline
    \end{tabular}
    \caption{Analysed software and tools.}
    \label{tab:analyzed-apps}
    \vspace{-12pt}
\end{table}

\begin{table*}[t]
    \footnotesize
    \centering
    \setlength{\tabcolsep}{3pt}
    \begin{tabular}{|l|l|l|l|l|Hc|l|}
\hline

                                      &                     &              & {\bf Mis-}                  &             & & {\bf Attacker}   &         \\
{\bf Section -}                             &                     &              & {\bf interpretation} &             & input 	  & {\bf can choose}                   &         \\
{\bf Category} 	                          & {\bf DNS use-case}        & {\bf Software}     & {\bf is in}        & {\bf Record type} & validated \hspace{-5pt}  & {\bf domain}                   & {\bf Possible outcome(s)} \\ \hline

\ref{sec:app-resolver} - DNS          & Address-lookup      & Verisign Public DNS$^{(*)}$ & Resolver          & CNAME       & no & yes                      & Cache injection \\ \hline
\ref{sec:app-eduroam} - Eduroam	      & Service discovery	& radsecproxy  & Application       & NAPTR, SRV  & yes	& yes	                   & Strip TLS, hijack connection, Crash \\ \hline
\ref{sec:app-ldap} - LDAP	          & Service discovery	& openjdk	   & Application       & SRV	     & yes   &  \new{no}                    & Crash                               \\ \hline
\ref{sec:app-smtp} - Email	          & Authorization	    & libspf2      & Application       & TXT (SPF)   & yes & yes	                   & Crash, (potential code execution)   \\ \hline
\ref{sec:app-admin} - Admin tools     & Address-, Reverse-lookup	    & ping,openssh,trace	       & Stub resolver     & CNAME	     & n/a     & no                       & Terminal Escape Code injection          \\ \hline
\ref{sec:app-admin} - Web-interface   & Reverse-lookup	    & OpenWRT luci & Stub resolver     & PTR	     & yes  & yes	                   & XSS in Admin web-interface          \\ \hline
    \end{tabular}
    (*) Recursive service operated by Verisign at the time the research was conducted.
    \caption{Applications' categories with vulnerabilities and attacks exploiting them. }
    \label{tab:vulnerabilities}
    \vspace{-12pt}
\end{table*}

\subsection{DNS Caches}
\label{sec:app-resolver}

The attacks exploit the fact that domains and hostnames are not restricted to characters, and implements misinterpretation of domain names due to presence of \stt{"\."} and of \stt{"\000"} characters. These characters cause the appearance of \stt{"."} to be altered hence manipulating the subdomains of a given parent domain.

The attacker can trigger a DNS query directly when launching the attack against open resolver or can initiate the attack via an application which uses the target DNS resolver, e.g., a web browser or an Email server.

\subsubsection{DNS Cache Poisoning Attacks}\label{fig:injectdot}\label{fig:inject000}
In this section we present two types of cache-injection attacks which are based on domain name misinterpretation and verify them against popular DNS resolvers' software as well as against 3M open DNS resolvers in the Internet.
 We also show how to extend our poisoning attacks against forwarders and provide an example of the poisoning attack we launched using the Verisign Public DNS\footnote{Verisign Public DNS was operated by Verisign at the time the research was conducted. Neustar acquired the IP addresses from Verisign last November to be incorporated into its own UltraDNS Public service \cite{neustar}.} resolver.

$\bullet$ {\bf Attack \#1: Period injection.}
To inject a malicious DNS record or to overwrite a cached DNS record with a new value (controlled by an attacker), we design the following record set inject\textsubscript{\textbackslash.}: {\small {\tt www\textbackslash.target.com. A 6.6.6.6}}.

This attack requires the attacker to control a specially-malformed domain \stt{www\.target.com.} under the same parent domain (in this example \stt{com.}) as the domain of its victim, say \stt{www.target.com}. Since most client software %
does not allow triggering a query for a domain \stt{www\.target.com} directly, to perform injection of a malicious record into the victim's cache, the attacker can set up a CNAME record with arbitrary subdomain, e.g., \stt{injectdot.attacker.com}, as follows:

\vspace{-5pt}
{\scriptsize
\begin{verbatim}
    injectdot.attacker.com. CNAME www\.target.com.
    www\.target.com. A 6.6.6.6
\end{verbatim}
}
\vspace{-5pt}

When decoding these records naively without escaping the period (\stt{"\."}) it appears that \stt{www.target.com} has IP address \stt{6.6.6.6}. Caching this misinterpreted record after decoding leads to DNS cache injection.

$\bullet$  {\bf Attack \#2: Zero-byte injection.} We design the following record set inject\textsubscript{\textbackslash000}, which indicates end of data, for performing DNS cache poisoning.

\vspace{-5pt}
{\scriptsize
\begin{verbatim}
    injectzero.attacker.com CNAME
        www.target.com\000.attacker.com
    www.target.com\000.attacker.com A 6.6.6.6
\end{verbatim}
}
\vspace{-5pt}

When naively decoded and fed into a victim cache this record enables an attacker to inject records for arbitrary domains into the cache.
In this attack we also use a CNAME alias mapped to some secondary domain \stt{injectzero.attacker.com}, since triggering a query to \stt{www.target.com\000.attacker.com} without direct access to the resolver is not possible with most client software. 
When decoding this record set into a C-string without escaping the zero-byte after \stt{www.target.com}, the \stt{.attacker.com} is removed since it is after the end of data \stt{\000} value, the DNS software misinterprets the record and caches a record mapping \stt{www.target.com} to IP address \stt{6.6.6.6}. 

\subsubsection{Evaluation of the Attacks}
Every application-level DNS-cache running on a system which misinterprets the inject\textsubscript{\textbackslash.} or inject\textsubscript{\textbackslash000} payloads (See Table~\ref{tab:resolvertests}) is vulnerable to these attacks. In our Internet study (see Section~\ref{sec:resolverstudy}) we found that 105,854 open DNS resolvers (or 8\% of 1,3M) are vulnerable to our attacks. Our attack evaluation was automated hence did not include potentially vulnerable resolvers which could result in a successful attack when the evaluation was manually tailored per resolver. These cases include lost packets (we sent only one response to avoid loading the network), resolvers with multiple caches (the attack was tested once against each client-exposed-IP of the resolver). Adjusting our attack to these cases is straightforward, would however generate much more traffic to the tested systems. %

\subsubsection{No Countermeasures Against Cache Poisoning}\label{sc:res:nocountermeasures}
Classic countermeasures against DNS cache poisoning do not mitigate our cache poisoning attacks. The situation is even more risky when the same host is configured as nameserver and resolver, \cite{shulman2015towards}, as a lack of validation by the DNS resolver can allow the attacker to also manipulate the zonefile which is hosted on the same machine.

{\bf Defences against off-path attackers.} Defences against off-path attackers, such as [RFC5452] \cite{rfc5452}, are not effective against our attacks: we do not send the malicious DNS responses from spoofed IP addresses but respond from a nameserver that we control. Hence in our attacks the attacker does not need to guess the randomisation values, such as UDP source port and the TXID. The bailiwick check \cite{rfc2181}, which prevents the attackers from responding with values not under their domains is also ineffective against our attacks since the bailiwick check is applied over the records {\em before} the misinterpretation occurs.

{\bf Defences against on-path attackers.} Cryptographic defences, most notably DNSSEC [RFC4033-RFC4035], can not prevent our cache poisoning attacks \new{in common setups: in situations where upstream resolvers are used \new{the misinterpreted records are not detected by the downstream DNS forwarders, since those typically do not perform DNSSEC validation}\footnote{Neither dnsmasq, systemd-resolved nor OpenWRT or Fritz!Box SOHO routers perform DNSSEC validation by default.}. DNSSEC validation is performed by the recursive resolvers over the DNS records in line-format, \textit{before} the decoding and the misinterpretation occur. After the records successfully pass DNSSEC validation, they are cached in a ``misinterpreted'' form.}

\begin{figure}
     \centering
     \includegraphics[width=0.46\textwidth]{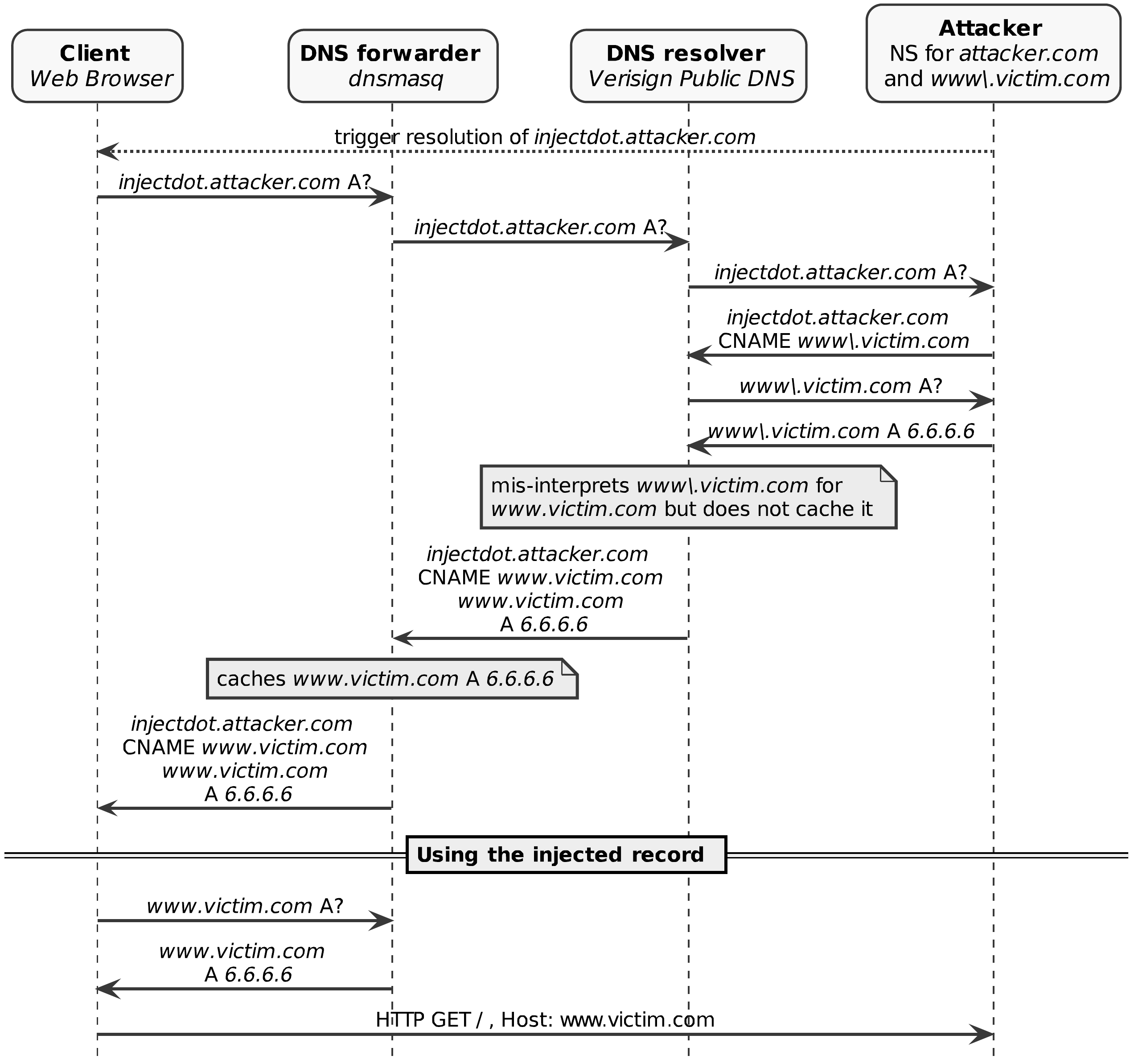}
     \caption{Downstream forwarder attack using dnsmasq with Verisign Public DNS misinterpretation of inject\textsubscript{\textbackslash{}.} payload.}
     \label{fig:injectdot-resolver-attack}
 \end{figure}

\new{{\bf Cross-zone CNAME caching.} Additionally to the misinterpretation, these attacks require resolvers to cache and process CNAME records across zone-boundaries, i.e., the resolvers must use the misinterpreted second record {\small\path{www.target.com\000.attacker.com}} from zone {\small\path{attacker.com}} to answer queries for {\small\path{www.target.com}}. While this is not typically the case for recursive resolvers, we validated such behaviour in dnsmasq, the most frequently used forwarder on our open resolver dataset: given the records
{\small\path{injectdot.attacker.com CNAME www.victim.com}} and  {\small\path{www.victim.com A 6.6.6.6}} in response to a query for {\small\path{injectdot.attacker.com}}, dnsmasq will answer queries for {\small\path{www.victim.com}} with \path{6.6.6.6}. We illustrate how this leads to a vulnerable configuration of forwarder and recursive resolver in the case dnsmasq is combined with a misinterpreting recursive resolver like Verisign Public DNS in Figure~\ref{fig:injectdot-resolver-attack}; our resolver evaluation is in Section~\ref{sec:resolverstudy}.}

\subsubsection{\new{Required attacker capabilities}} %
\new{To launch the inject\textsubscript{\textbackslash000} attack, the adversary only needs to control a nameserver for an arbitrary domain in the internet. In contrast, in order to launch the inject\textsubscript{\textbackslash.} attack the adversary has to control a specially crafted malicious sub-domain under the same parent-domain (e.g., {\tt com.}) as his target. This means that conducting this attack requires registering a sub-domain like {\tt www\textbackslash.target} via a domain registry. Applicability of this attack depends on ability of the attacker to register such sub-domains. 
For instance, a registry.pw for .pw reported that registering domain \stt{www\.asd.pw} was possible, while \stt{www.asd.pw} or \stt{asd.pw} was not (indicating that they are existing registered domains). Namely the attacker can register \stt{www\.asd.pw}  and use it to attack the existing victim domain \stt{asd.pw}.} %

\subsection{Eduroam Peer Discovery}
\label{sec:app-eduroam}

Eduroam federation uses Remote Authentication Dial-In User Service (Radius) \cite{RFC2058} for authentication of guest access. Radsecproxy is an application that implements Radius transport over TCP and TLS as well as dynamic peer discovery for servers which do not support these features themselves. Radsecproxy uses a shell-script-based method for dynamically updating the configuration to support the DNS lookups needed for Dynamic Peer discovery.
\new{The script is invoked with the domain component of the user's network access identifier (i.e., {\small\path{example.com}} in Figure~\ref{fig:radius-dynamic}) as its first argument by the radsecproxy server and outputs a new dynamic radsecproxy configuration for the user's realm.}
Example output of this script (called naptr-eduroam.sh) when invoked from shell is below: %

{\scriptsize
\begin{verbatim}
$ ./naptr-eduroam.sh example.com
server dynamic_radsec.example.com {
        host radius1.example.com:2083
        host radius2.example.com:2083
        type TLS
}
\end{verbatim}
}

\begin{table*}
    \footnotesize
    \centering
    \setlength\tabcolsep{4pt}
    \let\oldtt\tt
    \renewcommand{\tt}{\oldtt\scriptsize}
    \begin{tabular}{|l|l|l|l|l|l|}
    \hline
           & {\bf Variable}        & {\bf Record} &                                                        &                                          &         \\
        \# & {\bf in script}       & {\bf type}   & {\bf Malicious record data ({\tt dig}-escaped)}                    & {\bf Induced behaviour}                   & {\bf Outcome} \\ \hline
        1  & \$HOST    & NAPTR & \verb!\@6.6.6.6.!                                       & change dig DNS resolver                  & verification of vulnerability \\ \hline
        2  & \$HOST    & NAPTR & \verb!-f/some/file.!                                   & pass {\tt /some/file} as dig batch-file & disclose contents of {\tt /some/file}  \\ \hline
        3  & \$SRVHOST & SRV & \verb!asd\\n\\tinclude\\t/dev/zero\\n.!                 & read {\tt /dev/zero} as config file      & 100\% CPU utilisation         \\ \hline
        4  & \$SRVHOST & SRV & \makecell[l]{
\tt as.d\textbackslash{}\textbackslash{}n\textbackslash{}\textbackslash{}tmatchcertificateattribute\textbackslash{}\textbackslash{}t \\
\tt CN:/\textbackslash{}(.*+++++++++++++++++++\textbackslash{}(\textbackslash{}\textbackslash{}\textbackslash{}\textbackslash{}w+\textbackslash{})\textbackslash{}) \\
\tt /im\textbackslash{}\textbackslash{}n\textbackslash{}\textbackslash{}ttype\textbackslash{}\textbackslash{}ttls\textbackslash{}\textbackslash{}n\}\textbackslash{}\textbackslash{}n\%\%p. }
                                                                                       & provide malicious regex to regcomp()     & radsecproxy crash        \\ \hline
        5  & \$SRVHOST & SRV & \makecell[l]{
\tt 6.6.6.6\textbackslash{}\textbackslash{}n\textbackslash{}\textbackslash{}ttype\textbackslash{}\textbackslash{}tTCP\textbackslash{}\textbackslash{}n \\
\tt \textbackslash{}\textbackslash{}tsecret\textbackslash{}\textbackslash{}tsomething\textbackslash{}\textbackslash{}n\}\textbackslash{}\textbackslash{}n\%\%p. }
                 
                                                                                       & \makecell[l]{provide own RADIUS server \\
                                                                                         and disable TLS-authentication }             & unauthorised network access        \\ \hline
    \end{tabular}
    \caption{Radsecproxy exploits. The exploits were successfully verified in the lab and against large operators of Eduroam.}
    \label{tab:radsecproxy-exploits}
    \vspace{-18pt}
\end{table*}

\subsubsection{Radius Dynamic Peer Discovery}
\label{sc:appendix:radius}

\begin{figure}
    \centering
    \includegraphics[width=0.46\textwidth]{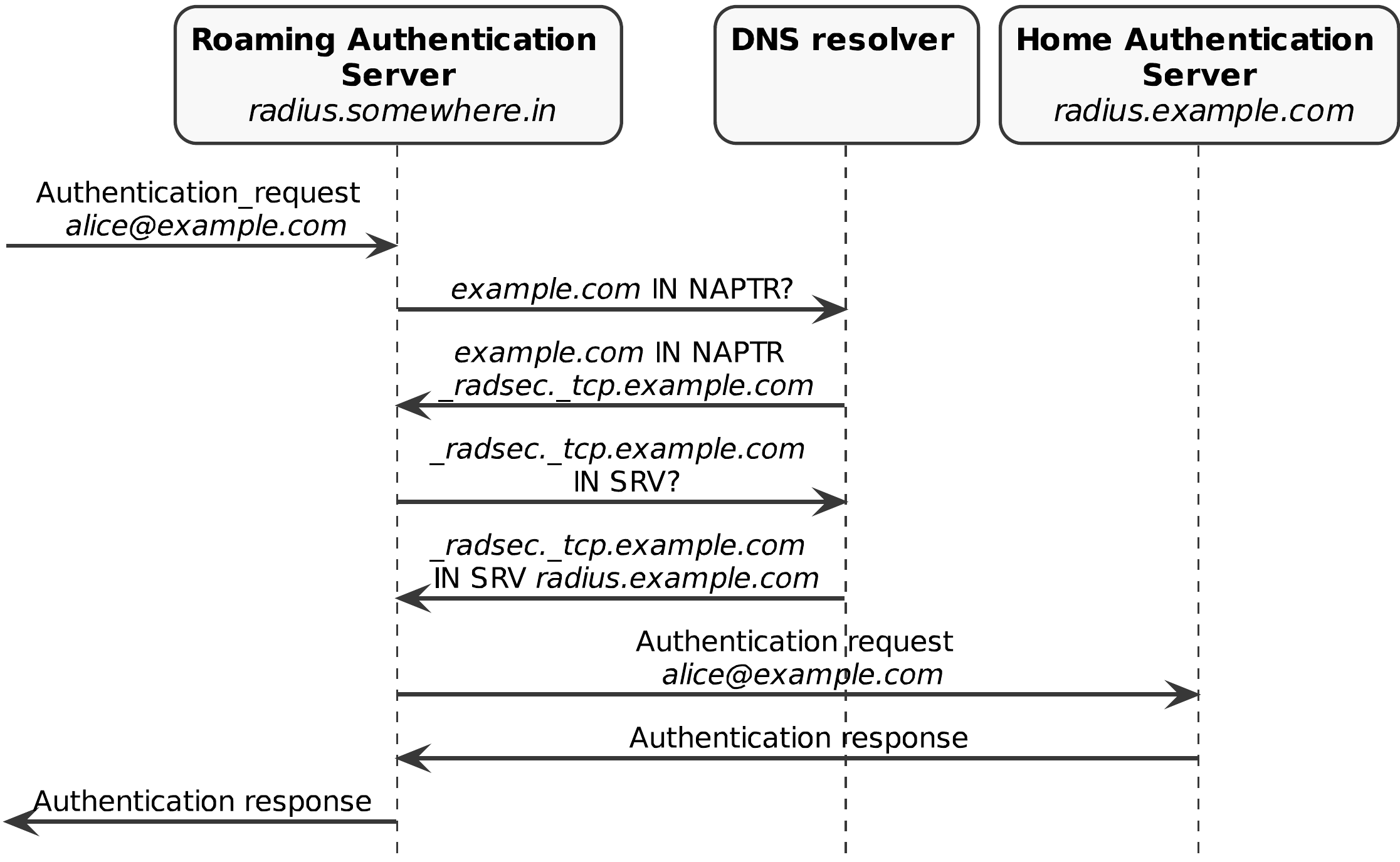}
    \vspace{-12pt}
    \caption{Radius Dynamic Peer discovery.}
    \label{fig:radius-dynamic}
    \vspace{-12pt}
\end{figure}

In this section we provide a detailed explanation of the radius dynamic peer discovery process illustrated in Figure~\ref{fig:radius-dynamic}, as well as how an adversary can abuse the mechanism to trigger queries.
First, a client (alice) connects to a wireless access point at the campus of \stt{somewhere.in} (the domain of that university), providing authentication material including her network access identifier (NAI) \stt{alice@example.com}. The access point then forwards the authentication request to the roaming authentication server at \stt{somewhere.in}. From alice's NAI this server defers that alice's authentication request needs to be routed to the home authentication server of \stt{example.com}. To find the home authentication server, it issues DNS queries for \stt{example.com} \stt{IN NAPTR?} followed by an \stt{SRV} query of the domains listed in the \stt{NAPTR} record. Finally, the roaming authentication server forwards the authentication request to the home authentication server at \stt{radius.example.com}, which answers the request.
The attacker sets up his own domain \stt{attacker.com} and configures his nameserver to answer with one of the attack payloads from Table~\ref{tab:radsecproxy-exploits}. The attacker provides a username \stt{user@attacker.com} when connecting. This leads the roaming authentication server to send DNS requests via its resolver to the attacker's nameserver. Depending on the payload, a corresponding attack is launched against the roaming authentication server.

\subsubsection{Attacks Against Radsecproxy}
\label{naptr-eduroam.sh-vulnerabilities}\label{sec:radsecproxy-exploits}
We found multiple security vulnerabilities in the script \cite{naptr-eduroam} for Dynamic Peer Discovery in eduroam, see vulnerabilities in Table~\ref{tab:radsecproxy-exploits}. The vulnerabilities allow an attacker to control various variables inside the script as well as in the generated dynamic configuration. These vulnerabilities are caused by the lack of input validation of the resulting output of \stt{dig} as well as the usage of \stt{printf}, which negates the escaping of special characters done by \stt{dig}. %

To initiate an attack the attacker causes the target system to issue a query to its domain. To make radsecproxy query for a domain of attacker's choice, the attacker just needs to attempt to log-in at an eduroam access point with a username ending with the malicious domain. This triggers {\tt NAPTR} and {\tt SRV} queries to locate the correct authentication server, see messages exchange in  Figure \ref{fig:radius-dynamic}.

\paragraph{No validation of {\tt dig} output:}~By changing the NAPTR record's replacement field ({\tt\small \$HOST}), the attacker can control
one argument to \stt{dig}, which is not checked for its format. We exploited this to launch two attacks: 

$\bullet$  {\bf Attack \#1 in Table \ref{tab:radsecproxy-exploits}.} Make dig query an attacker-chosen DNS resolver, instead of the default resolver from \stt{/etc/resolv.conf}. We used this attack to verify the vulnerability remotely without causing damage to the tested eduroam network. 

$\bullet$  {\bf Attack \#2 in Table \ref{tab:radsecproxy-exploits}.} Make dig use {\em any file} on the radsecproxy-system as a ``batch-file'' (option \stt{-f}), thereby querying all lines in this file as DNS queries.
 Attackers which are located on-path to the DNS resolver can apply the second attack to read arbitrary files from the system.

\paragraph{Vulnerable usage of {\tt printf}:}~The (double) use of printf in {\small\path{naptr-eduroam.sh}} allows the attacker to inject arbitrary strings into the dynamically generated configuration file via a format-string attack. The reason is that printf removes the escaping done by dig over the user input, which is given in format specifier argument. This allows the attacker to access radsecproxy's configuration parser and subsequent \stt{confserver_cb} function, which can be used to make the process read any file on the fileystem using
\stt{include /path/file}. 

$\bullet$  {\bf Attack \#3 in Table \ref{tab:radsecproxy-exploits}.} We evaluated a `resource starvation' attack using \stt{/dev/zero} as an input, and caused an infinite 100\% CPU usage loop in the configuration file parser. %

$\bullet$  {\bf Attack \#4 in Table \ref{tab:radsecproxy-exploits}.} In this attack we demonstrate how the attacker can manipulate the generated dynamic configuration file, specifying a TLS certificate CommonName (CN) regular expression. This expression is passed to the libc's \stt{regcomp()} function, which on many implementations of libc (e.g., glibc) has known, unfixed vulnerabilities\footnote{E.g., \cite{cve-2010-4052} can still be
exploited on current Ubuntu and used in attack \#4.}
which can be used, e.g., to crash radsecproxy via stack consumption.

$\bullet$  {\bf Attack \#5 in Table \ref{tab:radsecproxy-exploits}.} When the attacker provides a functional server configuration it can also override parameters of the dynamically generated server entry, most importantly the \stt{type} parameter. When changing the \stt{type} parameter to TCP and providing a known \stt{secret}, the attacker can make radsecproxy connect to his own radius server despite not having a trusted TLS certificate from the eduroam-PKI. 
Attack \#5 can be used to allow or deny access or log any authentication attempt for users using the attacker's domain as a realm. 
This enables the attacker to use any eduroam network effectively unauthenticated. In our evaluations we exploited this attack to even successfully inject malicious authentication server of the attacker for third-party domains, which enables the attacker to log usernames and/or hashed credentials when the wireless clients fail to verify the TLS-certificate provided in the protected-EAP tunnel to the attacker's RADIUS server.

\subsubsection{Evaluation of the Attacks}
\label{sec:radsecproxy-verify}

All listed exploits and outcomes were verified in the lab using the latest version of radsecproxy. We also validated real-world applicability of the attacks on different eduroam networks (of two research institutions and university) by exploiting the vulnerabilities listed in this section. We launched exploit \#1 against large operators of eduroam infrastructure. This exploit causes no harm but demonstrates that the infrastructure is vulnerable and uses the \stt{naptr-eduroam.sh} script.

\subsection{LDAP Peer Discovery}
\label{sec:app-ldap}

To locate the appropriate LDAP server dynamically, an LDAP client supporting dynamic peer discovery extracts the domain components, re-creates the domain name (e.g., {\small\path{example.com}}) and queries the DNS SRV-record for {\small\path{_ldap._tcp.example.com}}. This query is triggered either at application startup or when a user tries to connect to the LDAP-using service.
In addition to SRV lookups, LDAP also supports the URL-based description of search operations \cite{RFC4516}. For example, a URL for a search operation for john's user account entry may look like {\small\path{ldap://ldap.example.com:389/uid=john,gid=users,dc=example,dc=com}}. This instructs the LDAP client to connect to the LDAP server at {\small\path{ldap.example.com}}, port 389 and look for an entry with Distinguished Names (DN) {\small\path{uid=john,gid=users,dc=example,dc=com}}.
The attacker triggers queries by attempting to connect to the LDAP-using service.

\subsubsection{LDAP Injection Attacks}
When the SRV lookup is used in combination with LDAP URLs, it opens an attack vector which is caused by the SRV lookup handling of LDAP client implementations: %
the LDAP URL is checked for a hostname, and if it is not present, the hostname is looked up using SRV requests and pasted into the existing LDAP URL. An attacker controlling the SRV record can inject arbitrary characters into the URL, changing the URL path component and thereby the requested resource's DN or filter expression.

Algorithm~\ref{alg:ldabsrv} shows the LDAP peer discovery process as implemented by OpenJDK. The function \textit{ConnectURL} is called with an LDAP URL like \stt{ldap:///uid=john,gid=users,dc=example,dc=com} and parsed into a URL. The URL is then tested whether it includes a hostname, and if not the domain component (\stt{dc=}) parts of the LDAP distinguished name (DN) are used to construct the the query domain for dynamic peer discovery. In our case this is the domain \stt{example.com}, so the process continues by requesting that domains LDAP \stt{SRV} record at \stt{_ldap._tcp.example.com}. The hostname and port included in this SRV record are now used to construct a new LDAP URL by concatenating the hostname and port with the part of the path of the old LDAP URL. Finally, the new URL is used to call \textit{ConnectURL} again, this time taking the other path and connection to the LDAP server at the specified hostname.

We show how the user input concatenated to an LDAP query can change the meaning of the query by injecting control characters like braces, similar to SQL injections. %
Our LDAP injection uses the contents of the SRV record for dynamic peer discovery (instead of direct user input) and leads to information disclosure, or authentication as a different user.

\begin{algorithm}[t]
{\scriptsize
 \SetKwProg{Fn}{Function}{ is}{end}
 \Fn{ConnectURL(ldapurl: URL)}{
  \uIf{ldapurl.host == None }{
   domain = extractDC(ldapurl.path) \\
   hostname,port = lookupSRV(domain) \\
   ldapurl = new URL("ldap://" + hostname + ":" + port + "/" + ldapurl.path) \\
   ConnectURL(ldapurl)
  }
  \Else{
   ip = lookupA(ldapurl.host) \\
   // Proceed with connection ... \\
  }
 }}
 \vspace{5pt}
 \caption{LDAP SRV lookup.}
 \label{alg:ldabsrv}
 \vspace{-10pt}
\end{algorithm}

$\bullet$  {\bf Attack \#1: Privileges escalation.} When executing Algorithm \ref{alg:ldabsrv} with the following URL
{\small\path{ldap:///uid=john,gid=users,dc=example,dc=com}} and the SRV record set to
\vspace{-5pt}
{\small\begin{verbatim}
  _ldap._tcp.example.com IN SRV ldap.example.com/
  uid=admin,gid=users,dc=example,dc=com????.
\end{verbatim}}
\noindent the resulting URL becomes {\small\path{ldap://ldap.example.com/uid=admin,gid=users,dc=example,dc=com????./uid=john,gid=users,dc=example,dc=com}},
which means the client will search for user admin instead of john, enabling john to execute actions with admin privileges.
This attack enables to circumvent security mechanisms like LDAP over TLS
({\small\path{ldaps://}}) because it changes the information in the URL before it is transmitted over the TLS secured channel.

$\bullet$  {\bf Attack \#2: Denial-of-Service via malformed records.}
The LDAP SRV lookup function calls itself recursively after looking-up an SRV record, see Algorithm \ref{alg:ldabsrv}. 
We manipulate an SRV record so that the resulting URL does not contain a hostname-component, which then causes an infinite recursion and crashes the `LDAP-using' application with a stack overflow. 
\subsubsection{Evaluation of the Attacks}\label{fig:ldapopenjdk-crash}
We tested attack \#1 against two LDAP library implementations (ldapsearch and opejdk's javax.naming). We find that both applications use a potentially vulnerable LDAP peer discovery algorithm, which just concatenates the SRV record with the rest of the URL and do not check the contents of the SRV record for sanity. However, in both implementations, the DN (e.g., {\small\path{uid=john,gid=users}}) from the LDAP URL is actually ignored and must be given in an additional function call or parameter in order to allow execution of multiple search queries after the connection to the server has been established. %
We verified attack \#2 experimentally using the record we constructed, and evaluated it against openjdk's 11.0.6 javax.naming API: 

\vspace{-6pt}
{\scriptsize\begin{verbatim}
  _ldap._tcp.attacker.com. IN SRV /dc=attacker,dc=com.
\end{verbatim}}
\vspace{-6pt}

\textbf{\new{Triggering a query.}}
\new{A query for the LDAP \stt{SRV} record is triggered when a new connection to the LDAP server is created, i.e., when a user triggers an action which requires an LDAP-lookup such as logging into a web application which uses LDAP for user management.}
However, to execute the attack, the attacker must either be able to (1) control the full LDAP DN or (2) modify the SRV record on the network via a MitM position. \new{In our evaluation we tested the implementation of LDAP middleware/libraries, which do not restrict how the LDAP DN is set. However, typical applications will restrict control over the LDAP DN to the components relevant for the user\footnote{\url{https://docs.spring.io/spring-ldap/docs/current/reference/}}, such that control over the necessary \path{dc=} components is not available to the attacker.}

\subsection{Domain-Based Anti-Spam Validation}
\label{sec:app-smtp}

Sender Policy Framework (SPF) \cite{rfc7208} is a domain-based mechanism to prevent forgery of SMTP envelope headers. To trigger an SPF DNS query, the attacker needs to send an Email to an SPF-supporting Email server.

\begin{figure}
    \centering
    \includegraphics[width=0.46\textwidth]{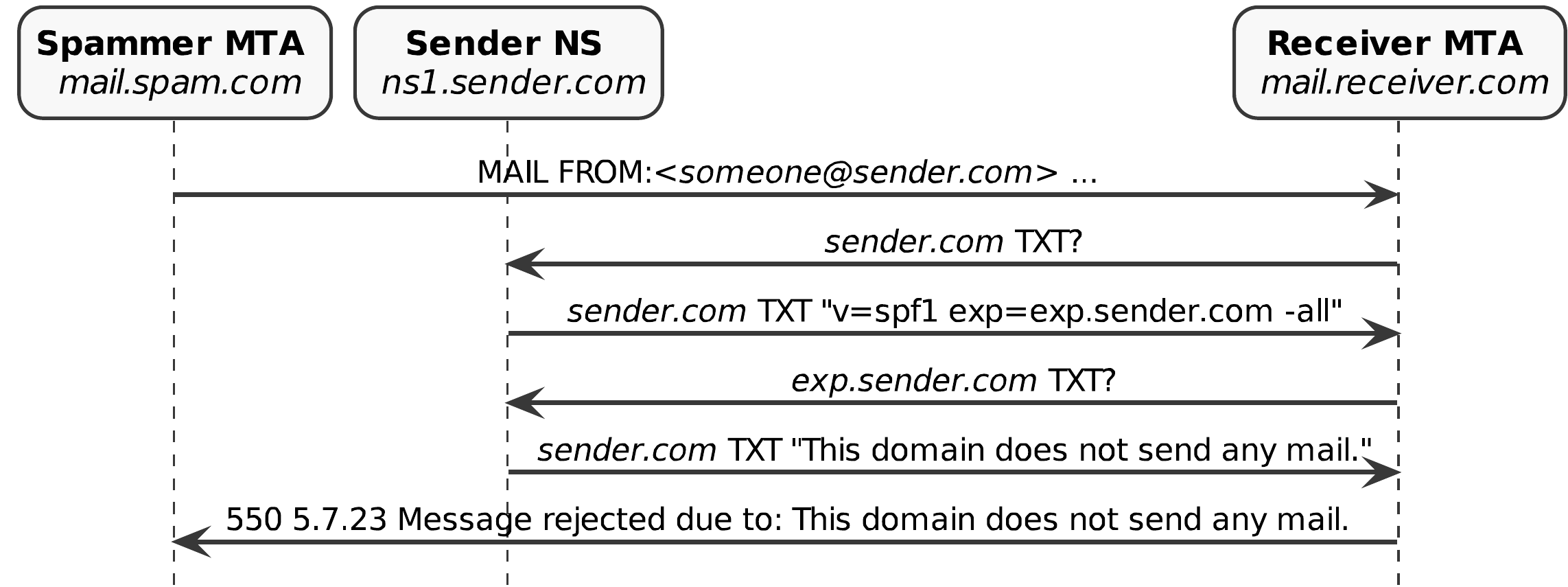}
    \caption{\new{\texttt{SPF} resolution example.}}
    \label{fig:spf-example}
\end{figure}

We provide a detailed explanation of the Email \stt{SPF} validation process shown in Figure~\ref{fig:spf-example} in the case where an incoming Email is rejected: first, an non-authorised Email transfer agent at \stt{mail.spam.com} (Spammer MTA) connects to the mail server at the receiver domain (\stt{mail.receiver.com}) and tries to send a Mail coming from \stt{someone@sender.com} to a mailbox at \stt{receiver.com} using SMTP. To check if the Spammer MTA is authorised to send mail from \stt{sender.com}, the Receiver MTA will query the DNS for the SPF records for \stt{sender.com}. In this case the record indicates that no one is authorised to send mail from that domain (option -all) and that a detailed explanation why the mail is rejected is stored at \stt{exp.sender.com}. After the receiver MTA has received this record it will decide to reject the Email and fetch the explanation from \stt{exp.sender.com} via DNS to include it together with the rejection message. Finally, is parses the rejection message, replaces any included macros and sends it back to the spammer MTA notifying it that the mail was rejected and why.

\subsubsection{Attacks Against Checks of SPF Records}
\label{sec:postfix-policyd}

$\bullet$ {\bf Attack \#1: Injection against policyd-spf.} The default implementation, e.g., Ubuntu \cite{postfix_spf_ubunutu} used by postfix, for checking SPF records is based on a separately running daemon called policyd-spf. This daemon is listening at a unix socket for responses to determine whether an Email should be rejected or not according to the SPF records \cite{postfix_policyd_docu}. The interface to this daemon is line based, the client (postfix) provides properties of the received Email line-by-line and submits the request with an empty line, as shown below. In this example policyd-spf session, the client's request lines are marked with `\textbf{\tt >}`, deamon response lines with `\textbf{\tt <}`.

\vspace{-6pt}
{\scriptsize
\begin{verbatim}
    > request=smtpd_access_policy 
    > protocol_state=RCPT  
    > client_address=192.168.234.20 
    > sender=someone@hardfail.example.com 
    > recipient=vagrant@postfix 
    ... 
    > policy_context=
    < action=550 5.7.23 Message rejected due to: SPF fail - 
    not authorized. Please see http://www.openspf.net/Why?s=helo;
    id=hardfail.example.com;ip=192.168.234.20;r=<UNKNOWN>
\end{verbatim}
}
\vspace{-6pt}

The server (policyd-spf) will answer with a single line providing information on how to proceed. We construct a malformed SPF record using the SPF {\tt exp=} parameter. This parameter allows to include an explanation why an Email was rejected. We inject control characters into this client-server interface by including them in the rejection message, specified as a separate DNS record. The attacker can further include newline characters ({\tt"\textbackslash{n}"})
to create additional output lines in the line-based connection to policyd-spf. This is interpreted as the responses to requests asked in the future thereby changing the SPF result for the next Email. 

$\bullet$  {\bf Attack \#2: Stack buffer overflow in libspf2.} Libspf2 is a library for checking SPF records for incoming Email messages for Mail Transfer Agents. Libspf2 is used by some versions of the command line utility {\small\path{spfquery}} and also directly from SMTP server source code. Because of the complexity of the library we used fuzz-testing with {\small\path{afl-fuzz}} against a custom-built application calling libspf2 functions. This allowed us to provide the SPF record as a file rather than via the network to test libspf2 against potential vulnerabilities exploitable via malformed SPF records. The evaluations showed that libspf2 is vulnerable to the malformed records attack, see malicious SPF record payload in Figure \ref{fig:libsfp2-payload}.
\begin{figure}
    \centering
{\scriptsize
\begin{verbatim}
           attacker.com TXT "v=spf1 exp=exp.attacker.com"
           exp.attacker.com TXT "AAAAAA..." ; (510 times)    
\end{verbatim}
}
\vspace{-12pt}
   \caption{\small{libsfp2 exploit with malicious SPF record payload.}} %
    \label{fig:libsfp2-payload}
    \vspace{-16pt}
\end{figure}
We performed attacks against the {\small\path{spfquery}} command line utility using the vulnerable record set which resulted in `stack-smashing detected' error and crashes, further allowing remote-code-execution. %
This attack exploits a stack-buffer overflow while parsing the SPF explanation macro. %

\subsubsection{Evaluation of the Attacks}
We evaluated attack \#1 against postfix using policyd-spf-perl, and were able to inject additional lines of output to the unix socket, showing that policyd-spf-perl does not verify the contents of the SPF explanation record. %
In contrast to other attacks in this work, attack \#2 cannot be prevented by validating the DNS records since the malicious SPF record presents a theoretically valid SPF explanation message.

\subsection{Administrative Tools}
\label{sec:app-admin}

In the attacks that we presented until now, the applications implemented the DNS lookup themselves, not by using an API like {\tt gethostbyname()}, where the behaviour is standardised and data validation is performed by the systems' stub resolvers. \new{In this section we present vulnearbilities in applications} which do not implement DNS lookups themselves \new{but use the system stub resolver to do so. First, we analyse different stub resolver implementations in Section~\ref{sec:recorprocessing} and then show vulnerabilities in applications using these stub-resolvers in Section~\ref{sc:app:admintools}.}

\subsubsection{DNS Record Processing in stub resolvers.}
\label{sec:recorprocessing}
Libc is the C standard library for C programming language, specified in ANSI C and is a subset of C library POSIX specification. There are different implementations of the standard C system library. We experimentally tested, as well as analysed the source code of, all the major implementations of the C system library and except two found them to be vulnerable, see Table \ref{tab:resolvertests}. We explain our analysis of the DNS record processing on two implementations: glibc\footnote{The GNU's project implementation of the C standard Library} and on uClibc\footnote{The Linux standard library for mobile and embedded devices.}. We selected those implementations as examples because they represent two distinct methodologies that we observed in processing DNS records. The other tested implementations are similar to uClibc. We demonstrate the processing on the {\small\path{gethostbyname()}} library function as an example; the same applies to other calls.%

\begin{figure}[t]
    \centering
    \includegraphics[width=0.40\textwidth]{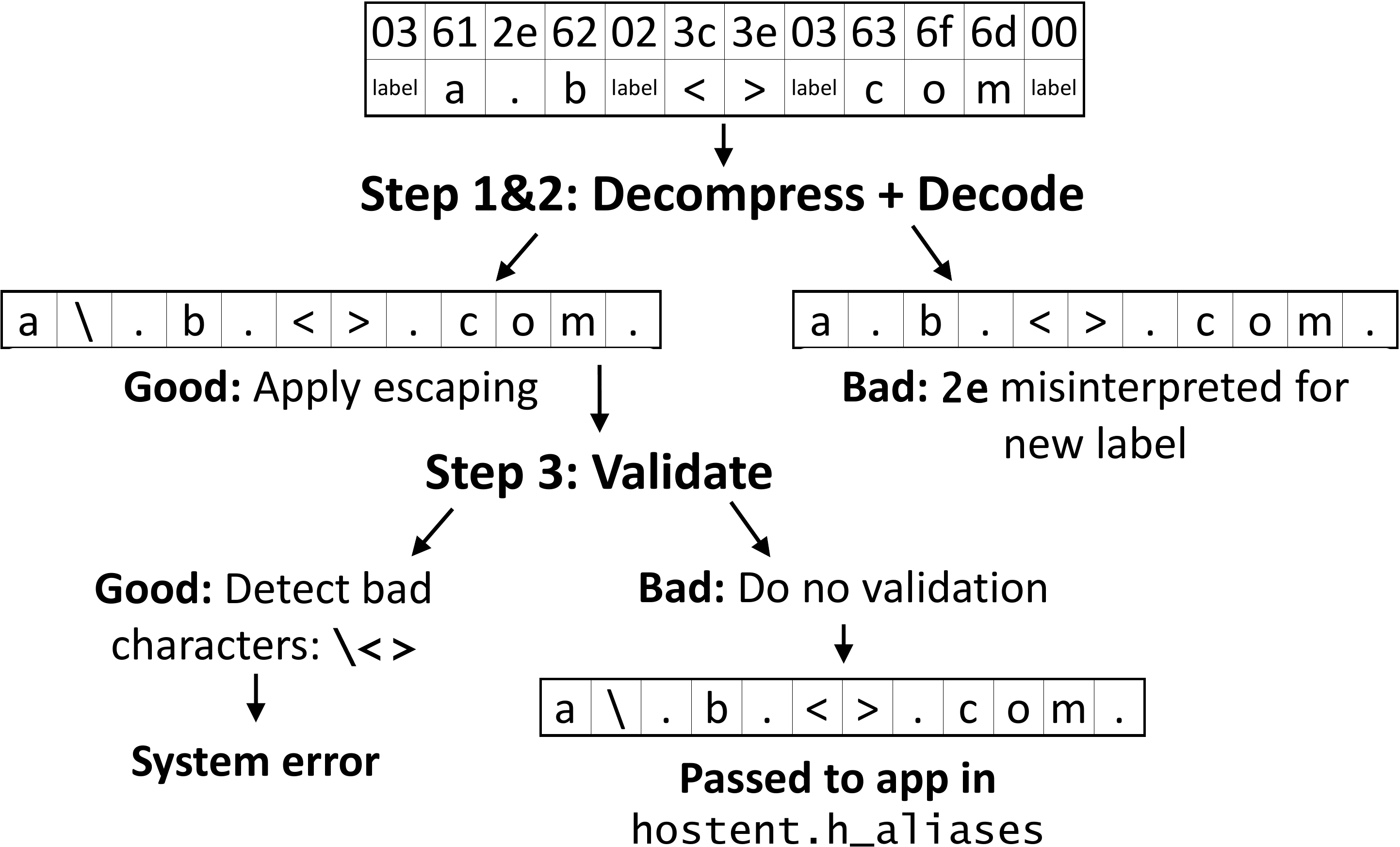}
    \caption{DNS record processing.}
    \vspace{-18pt}
    \label{fig:decode-flow}
\end{figure}

After a DNS response has been received by glibc (resp uClibc) library, it is first checked against the length field, DNS transaction identifier and the return code (e.g., OK, NXDOMAIN). The libraries then go through the resource record sets in the answer section and process each record. For each domain name in a DNS record the following steps are done, as shown with our example domain name \stt{036123...6d00} in Figure~\ref{fig:decode-flow}: (1) domain name compression is removed; (2) domain name is decoded from DNS line format into a (zero-terminated) string; (3) domain name is validated. We explain these steps next.

{\bf DNS decompression and decoding into a string.} These two steps are typically done simultaneously in one function, e.g., {\small{\tt dn\_expand}} for glibc, or {\small{\tt \_\_decode\_dotted}} for uClibc. When decoding a domain name into a string, the resolver must ensure that the characters which cannot be represented in an ASCII string, must be escaped appropriately [RFC4343]\cite{rfc4343}. This also applies to zero-bytes (which would otherwise be interpreted as string terminators) and period characters (which would otherwise be interpreted as label-separators). Escaping values outside the range of 0x21 (\stt{"!"}) to 0x7E (\stt{"~"}) is required by \cite{rfc4343}. 

In our example in Figure \ref{fig:decode-flow} this means that to avoid confusion with label separators the second byte (0x3e) of the first label must be expressed as \stt{"\."} instead of \stt{"."}. The final decoded domain name is then \stt{"a\.b.<>.com."} if decoded correctly applying escaping to non-printable characters, or \stt{"a.b.<>.com."} if decoded incorrectly when escaping is not applied. 

Our analysis shows that glibc applies escaping to the decoded domain name, while uClibc does not. This means that any record which contains zero-bytes or dots inside labels will be misinterpreted during decoding when processed by uClibc and other non-printable characters will be included in the returned string unescaped. Such incorrect decoding logic can allow cache-injection attacks when the misinterpreted record for the domain \stt{a.b.<>.com.} is cached and re-used in another context, as we show in Section~\ref{sec:app-resolver}.

{\bf Domain name validation.} After the record has been decoded, it should be validated to check that it represents a valid hostname. The POSIX standard defines {\small\path{getnameinfo()}} to return `hostnames', not `domain names'. Notice that domain names are defined in [RFC1034, RFC2181] \cite{rfc1034,rfc2181} as a list of binary labels that can contain any value. The only limitation over domain names is on the length of the name [RFC2181] \cite{rfc2181}: 63 octets per component, 255 octets for a domain name. Values like, brackets \stt{([, ])}, colons, NULs \stt{(\000)}, newlines, backslashes, and so on are all legal. In contrast, hostnames are only allowed to contain alphanumeric characters (\stt{"A-Z","a-z","0-9"}), hyphens (\stt{"-"}) and dots (\stt{"."}) to separate labels, \new{as defined in} [RFC952] \cite{rfc952}, \new{this specification is referenced by newer standards [RFC1123, RFC2181, RFC3492] \cite{rfc1123,rfc2181,rfc3492}}. As a result the system library should validate that all the returned values represent valid hostnames, not domains. Our evaluation shows that most stub resolvers use a naive domain name decoding logic which misinterprets {\small\path{"\."}} for {\small\path{"."}} and {\small\path{"\000"}} as a string delimiter; we show how to exploit this for cache poisoning attacks in Section \ref{sec:app-resolver}. 

In our example in Figure \ref{fig:decode-flow} step 2, we assume that the library has decoded the domain name correctly, hence the value \stt{a\.b.<>.com.} is passed on to the validation step. Based on the logic that the POSIX standard defines the return value of {\small\path{gethostbyname()}} as a hostname the library should ensure that the hostname does not contain invalid characters, and if it finds any, it should signal an error to the application. If this is not done, the string \stt{a\.b.<>.com.} is passed to the application unchanged and exposes to a range of vulnerabilities as shown in Sections~\ref{sec:app-admin}. Again we review the functions in glibc and uClibc and find that only glibc implements this validation correctly, uClibc does not validate the decoded domain name at all.

Notice that the steps 1-3 are always needed to transform a domain name from its line format to a (zero-terminated) ASCII string. An implementation might choose to switch the order of steps 2 (decoding) and 3 (validation), or combine these steps into one, but this does not change the fact that both steps are needed to correctly implement parsing of a line-format domain names into a hostname string. Therefore, our analysis of glibc and uClibc can be extended to other resolver implementations. We validated this assumption experimentally and via code review in popular system resolver implementations, and summarise the results in Table \ref{tab:resolvertests}. Our results indicate that most stub resolvers do not check that domain names constitute valid hostnames. The other libc implementations, such as \textit{dietlibc} and \textit{windows}, all result in the same incorrectly processed output. We show how to exploit lack of validation to attack applications which do not expect special characters inside hostnames, launching different injection attacks, such as XSS or ANSI terminal escape code injection. 

\subsubsection{Attacks Against Administrative Utilities}\label{sc:app:admintools}
We demonstrate attacks against two vulnerable example applications from different contexts: Windows 10's {\it ping} as an example of an simple command-line utility and {\it OpenWRT LuCi} as an example of a integrated web-based administration interface.

$\bullet$ {\bf Attack \#1: Windows ping.} Ping shows the CNAME alias of the ping-ed host without checking the alias for disallowed characters. We use this to inject arbitrary bytes into the output of ping, which is then displayed by the terminal. By including ANSI terminal escape codes in the CNAME record, an attacker controlling the DNS response can manipulate the terminal output by moving the cursor, replacing already printed data or changing the Window title\footnote{Terminal-escape injection vulnerabilities were found in applications like web-servers \cite{cve-2009-4487,cve-2013-1862}, however they were exploitable via direct input over HTTP.}.

  We tested a similar attack against linux ping, traceroute and openssh. Different than windows ping, these applications query the reverse PTR record of the remote address. We found that none of the applications perform input validation of the returned domain name and therefore allow the same kind of attack.%

$\bullet$ {\bf Attack \#2: Cross-site scripting (XSS) attack in OpenWRT LuCi.}
OpenWRT is an aftermarket operating system for residential-gateway routers based on linux. It uses uClibc as its standard C library. This means that any application included in OpenWRT which does not check CNAME or reverse-DNS responses itself is vulnerable to our attacks. %
We explore the web-based LuCi \cite{the_openwrt_project_luci_2020}
administration interface which can be used to configure all of a router's settings via a web-browser.

\begin{figure}[t]
    \centering
     \vspace{-8pt}
    \includegraphics[width=0.4\textwidth]{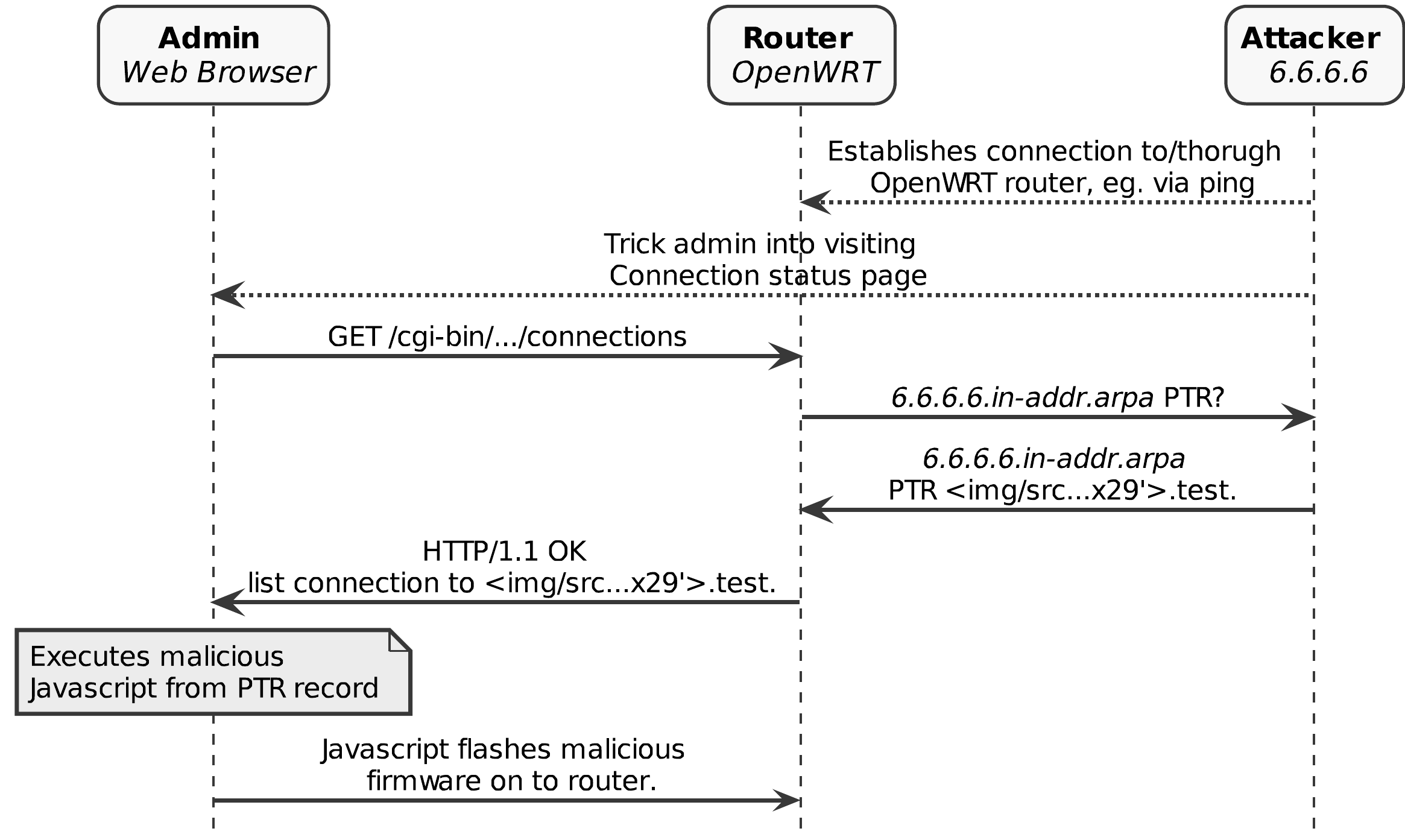}
    \vspace{-8pt}
    \caption{Attack flow against OpenWRT.}
    \vspace{-17pt}
    \label{fig:openwrt-attack}
\end{figure}

Apart from letting the user change the router's settings, LuCi also contains a status page listing all currently active connections through the router at {\small\path{/cgi-bin/luci/admin/status/realtime/connections}}. 
When a user opens this page, LuCi performs reverse-DNS lookups for all IP addresses currently connected to or through the router. The attacker abuses this to trigger reverse queries for its own IP addresses in the reverse-DNS tree and delivers malicious DNS records from there. The attacker also maintains a connection to or through the OpenWRT router when the victim opens the `Connections page', by continuously sending ICMP echo-request (`ping') messages to the router.
Using the {\small\path{6.6.6.6.in-addr.arpa}} record shown in Figure \ref{fig:cnameptrtestrecords} we were able to successfully launch this attack by placing javascript code inside the {\tt PTR} record of the attacker's IP address which is not validated by the LuCi web interface. This record is queried when the user views the `Connections page' and is placed in the page's HTML code which executes the malicious code. A successful attack allows different exploits, as an example we created a record which, when injected, loads a third party script from an external HTTP server. This script then issues various requests to the OpenWRT configuration interface in name of the user which finally results in replacing the OpenWRT firmware running on the device with a malicious attacker-provided firmware. We show an example of a full attack flow against OpenWRT using this vulnerability in Figure~\ref{fig:openwrt-attack}. %

\begin{table}[t]
    \centering
    \footnotesize
    \setlength{\tabcolsep}{3pt}
    \newcommand{\mr}[2]{\multirow{#1}{*}{\makecell{#2}}}
    \begin{tabular}{|lH|c|c|l|}
\hline
\mr{2}{Use-Case} & Type  & \mr{2}{Apps}           & vali-         & \mr{2}{attack prevented by} \\
          &              &                        & dates         & \\
\hline
Address   & Browser,     & Chrome, Firefox, Opera,& \mr{2}{no}    & Cache does not evaluate  \\
Lookup    & App-Cache    & Edge, unscd, java      &               & CNAME aliases for indexing \\
\hline
Service   & LDAP         & \mr{2}{ldapsearch}     & \mr{2}{no}    & Generated URL not used \\
discovery &              &                        &               & to specify LDAP query \\ \cline{3-5}
\hline
Authenti- & SMTP         & \mr{2}{policyd-spf + postfix} & \mr{2}{no} & postfix resets connection \\
cation    &              &                        &               & if second answer is injected \\
\hline
    \end{tabular}
    \caption{Non-validating applications without meaningful exploits.}
    \label{tab:faliedattacks}
\end{table}

\subsection{Impact of the Attacks}\label{sec:appsconclusions}

Our analysis of the attacks shows that none of the tested applications perform DNS input validation. Some  attacks do not result in meaningful exploits, see examples in Section \ref{sc:not:attack}, the causes are not defences against potentially malicious inputs, but rather not security related implementation decisions. We also showed attacks with limited impact, such as the ping exploit, to demonstrate that the problem (lack of DNS-input validation) is systematic and prevalent, affects different systems, services and tools and is not limited to isolated cases or a specific application. These examples show that developers do not check DNS results, which contain untrusted data. Furthermore, using such tools in certain scenarios could expose to the exploitable attack. For instance, output of utilities like Ping, can also be saved in a format which allows command injection, like an HTML report, or SQL database, this allows for a much more severe attack. %

\subsubsection{Non-validating applications without exploits}\label{sc:not:attack}
In this section we provide examples (listed in Table~\ref{tab:faliedattacks}) where DNS input validation vulnerability does not lead to meaningful attacks. We caution that the factors preventing meaningful attacks are not security related, but are due to different implementation considerations: (1) in browsers and app-caches, the cached records are only indexed by their query domain, not by additional domain aliases inside the response. (2) in ldapsearch, a potentially vulnerable peer discovery algorithm is used to generate an LDAP URL, but depending on the configuration this URL may not be used to specify the LDAP query, which would prevent the attacker from changing the query. (3) in policyd-spf if implementations are configured to check for additional data on unix socket, the result will be discarded to prevent desynchronisation with the policyd-spf daemon.

\section{Internet Evaluation on Open Resolvers}

\label{sec:resolverstudy}
In Section \ref{sec:appvulnerabilities} we examined how the popular DNS resolvers and stub resolver implementations built into operating systems and programming languages handle control characters in domain names and if they modify any of the maliciously crafted payloads needed to conduct our application-specific exploits. In this section we extend our evaluation to open DNS resolvers in the Internet, and confirm the results of our in-lab study in the Internet. Specifically, we evaluate behaviour when processing a set of crafted DNS records designed to trigger classic input validation vulnerabilities like SQL injection, as well as DNS-specific special cases like handling of period characters inside DNS labels. We do not evaluate attacks against the applications using the vulnerable resolvers, since this would result in an attack against an application or service in the network which we do not own. We do not risk evaluating even the `more benign' attacks, which when run against the servers that we setup do not cause critical damage. This is since our attacks can trigger unexpected outcome when evaluated against the servers in the Internet, e.g., due to differences in configurations. We observe this phenomena during the DNS cache poisoning evaluation of domains that we control against open resolvers in the Internet -- some of the DNS software which was found secure when running the cache poisoning attack against our servers, resulted in cache poisoning vulnerabilities in the Internet.

\subsection{Methodology}\label{sc:method}
We run the same set of queries as for the in-lab evaluation, but employ additional logging at the nameserver-level to gain knowledge about the nature of the resolvers we test. As this study targets the same class of resolver as our in-lab evaluation, the expected behaviour is the same as described in Section~\ref{sc:background}. %

{\bf Dataset.} To conduct the study, we use a dataset from Censys \cite{censys15} with 3M open resolvers.
We include baseline tests for each record type (A, CNAME, SRV, TXT) to ensure that the resolver supports the record type we use in our payloads and only consider resolvers who respond to our queries and return the correct result for all of these baseline tests. This results in 1,328,146 open resolvers from 228 different countries. 

Tests are conducted by using custom test applications which send DNS queries to the resolvers over the network or by calling the respective stub resolvers using the appropriate API, by calling the POSIX \stt{gethostbyname()} and \stt{getnameinfo()} functions.
To test handling of commonly used control characters like slash ({\tt "/"}), at ({\tt "@"}), zero-byte ({\tt "\textbackslash000"}), etc. we use the CNAME and PTR records with the injection payloads.
We furthermore test all the resolvers against the applications'-specific records listed in Section~\ref{sec:appvulnerabilities} whenever applicable. %
\new{All tested payloads are listed in Figure~\ref{fig:cnameptrtestrecords} in the order they appear in Table~\ref{tab:resolvertests} and \ref{tab:gethostbyaddr}.}

{\bf Forward-lookups.} We evaluate hostname-to-address records for all 3 groups of resolvers (resolver, stub and open resolver) by triggering a query to the domain name of the payload (e.g., {\small\path{cnameslash.example.com}}) and observing the response from the resolver. 

\new{{\bf Injection payloads.}} For the DNS-specific injection payloads, the test takes place in two stages: First we trigger a query to the domain name of the injection payload (i.e., {\small\path{injectzero.example.com}}) twice and observe \new{if} the result \new{was misinterpreted, ie. if the result CNAME is \path{www.target.com\000.example.com} (marked \xmark) or just \path{www.target.com} (marked \xmark$^5$)}. We then trigger a query to the potentially misinterpreted domain name (i.e., {\small\path{www.target.com}} instead of {\small\path{www.target.com\000.example.com}}) and observe if the IP address was successfully injected into the resolvers' cache \new{(marked \cmark or given in \% for open resolvers)}. We test each payload (inject\textbackslash000 and inject\textbackslash.) in both scenarios: via a CNAME-record which points to the malicious record (i.e., {\small\path{injectzero.attacker.com}}) and by triggering a query to the malicious domain (i.e., {\small\path{victim.com\000.attacker.com}}) directly. %

{\bf Reverse-lookups.} Reverse-DNS lookups (PTR) were tested against system stub resolvers only by setting the upstream DNS server to a custom controlled nameserver directly providing the records under {\small\path{in-addr.arpa}}-tree and triggering a reverse PTR lookup for the respective IP address, e.g., {\small\path{1.1.1.1.in-addr.arpa}} and observing the response. 

{\bf Additional considerations.} To enhance the robustness of our tests, we randomise all the queried domain names by prepending a random subdomain to ensure that the query is not cached before the test is conducted and processed by all the components of the DNS lookup chain. This also allows us to link the open resolver we sent the query to, to the final recursive resolver which connects back to our nameserver by matching these random prefixes. To prevent other users of the resolver to be negatively affected by our tests, we run these tests only against domains we own, which allows us to validate the full injection attack without performing an attack against any other domain.
\begin{figure}
    \centering
{\scriptsize
\begin{verbatim}
cnamebase.example.com  CNAME works.cnameslash.example.com
cnameslash.example.com CNAME t/t.cnameslash.example.com
cnameat.example.com    CNAME t\@t.cnameat.example.com
cnamexss.example.com   CNAME <img/src=''/onerror='alert
                             &#x28&#x22xss&#x22&#x29'>.
                             cnamexss.example.com
cnamesql.example.com   CNAME 'OR''=''--.cnamesql.example.com
cnameansi.example.com  CNAME \027[31\;1\;4mHello\027[0m.
                             cnameansi.example.com
\end{verbatim}

\begin{verbatim}
injectdot.example.com. CNAME www\.target.com.
www\.target.com.           A 6.6.6.6
injectzero.example.com CNAME www.target.com\000.example.com
www.target.com\000.example.com A 6.6.6.6

_ldap._tcp.example.com.  IN SRV /dc=example,dc=com.
_radsec._tcp.example.com IN SRV 6.6.6.6\\n\\ttype\\tTCP\\n\\t
                                secret\\tsomething\\n}\\n%
exp.example.com          IN TXT "AAAAA..." (510 times)
\end{verbatim}

\begin{verbatim}
1.1.1.1.in-addr.arpa   PTR   works.test
2.2.2.2.in-addr.arpa   PTR   te/st.test
3.3.3.3.in-addr.arpa   PTR   te\@st.test
4.4.4.4.in-addr.arpa   PTR   t\000t.test
5.5.5.5.in-addr.arpa   PTR   t\.t.test
6.6.6.6.in-addr.arpa   PTR   <img/src=''/onerror='alert
                             &#x28&#x22xss&#x22&#x29'>.test
7.7.7.7.in-addr.arpa   PTR   'OR''=''--.test
8.8.8.8.in-addr.arpa   PTR   \027[31\;1\;4mHello\027[0m.test    
\end{verbatim}
}\vspace{-5pt}
   \caption{\small{Injection payloads based on CNAME and PTR records.} } %
    \label{fig:cnameptrtestrecords}
    \vspace{-15pt}
\end{figure}

\subsection{Evaluation Results}\label{sec:openresolvers}

We present the results for forward-lookups in Table~\ref{tab:resolvertests} and the results of reverse-lookups in Table~\ref{tab:gethostbyaddr}. For each test, ticks (\cmark) mark that the resolver is vulnerable to this kind of payload and crosses (\xmark) mark that the resolver is not vulnerable. Note however that depending on the test, `vulnerable` means that the resolver performs as expected and conforms to the DNS standard (in case of non-stub resolvers and special character tests like cnameslash) or that it misinterprets a domain name which allows for a cache-poisoning attack (in case of injection payloads, such as inject\textbackslash000). %
\new{For injection payloads, we only call a resolver vulnerable (\cmark) when the malicious IP address was cached, otherwise we use (\xmark$^5$) to show that the misinterpretation occurs, but can only exploited in conjunction with a caching downstream resolver like dnsmasq (See Section~\ref{sc:res:nocountermeasures}).}
We list the percentage and absolute number of open resolvers vulnerable to each payload in the bottom of Table~\ref{tab:resolvertests}. 

{\bf Transparent handling of DNS records.} The results from Internet evaluation of open resolvers correspond to our in-lab evaluation: Around 96\% of all tested open DNS resolvers are transparent for application-specific payloads in DNS records. 

Comparing the CNAME-based cnameslash and SRV-based LDAP and Eduroam payloads, we can observe that there is some difference of resolver behaviour in handling these records, even thought they are based on the same property (including non-standard characters in a domain name field). This can be seen as a confirmation that many resolvers indeed ignore the contents of record type field that is not directly needed for the DNS resolution (like SRV), but do not do so for records which are important to finish the lookup (like A or CNAME).

{\bf Cache injection vulnerabilities.} For the injection specific payloads inject\textbackslash. and inject\textbackslash000, the evaluation results also \new{mostly} match the expectation from the in-lab evaluation. \new{Like all resolver implementations except Verisign Public DNS, most Open DNS resolvers which do handle the injection payloads transparently and are not vulnerable to our injection payloads.}

Nevertheless, we found 1.3\% to 4.6\% of the open DNS resolvers to be vulnerable to a cache poisoning attack without any further requirements, \new{which was verified by querying for the maliciously injected record.}
Overall, 8.0\% (or 105,854) of the open resolvers are vulnerable to cache poisoning via {\em any} of the injection payloads. %
This result is alarming considering that the inject\textbackslash000 attack does not require any attacker resources other than control over an arbitrary nameserver in the internet.

{\bf Misinterpretation analysis.} In our Internet measurement we observed the following phenomenon: the resolvers respond with the non-misinterpreted value in return to the first query for a record with injection payload in our domain. However, subsequent queries for the same resource record (with the same domain name) which are responded from the cache (without issuing a query to our nameserver) can result in one of the two outcomes: (1) a misinterpreted value or (2) non-misinterpreted value. Nevertheless a subsequent validation of the cached record confirms that in both cases the injection was successful and the target cache stored a misinterpreted record.

{\bf Vulnerable DNS software in Internet.} We use \path{version.bind} special query to infer the implementation and version of the open resolvers. While most servers do not respond to these queries (in our study 65\%, see column `our study` in Table~\ref{tab:resolvertests}), we find that, depending on the implementation and attack type, even those exact versions we have found not vulnerable during our lab evaluation were found vulnerable during our evaluation of open resolvers in the Internet, e.g., 19.2\% of \new{vulnerable resolvers were} Bind resolvers. In these cases we find that the misinterpretation causing the vulnerability does not occur in the `visible' software implementations, but rather in one of the upstream (forwarding) resolvers in the resolution chain\new{, which we cannot identify}. 

In some cases upstream forwarders is a public DNS service, like Google, OpenDNS and Cloudflare. We infer this by mapping the IP addresses of the DNS queries received by our nameservers to the Autonomous System (AS) numbers of the resolvers sending the queries. %

{\bf [RFC1034] non-compliant resolvers.} 12\% of the open resolvers in the Internet exhibit the same flaws as we found in stub resolver implementations. These are misinterpretations when decoding line-format domain names into zero-terminated string. Not only this allows for cache injection attacks, but it also indicates that those resolvers do not follow the recommendations in [RFC1034] \cite{rfc1034}, requiring that the DNS software should store domain names in their line format and not as decoded strings.

\begin{table*}[ht]
    \scriptsize
    \centering
    \setlength{\tabcolsep}{1pt}
    \begin{tabular}{|c|l|c|cc|ccc|cc|cc|ccc|HHH} %
    \cline{1-15} %
    & DNS Payload shown in                   & \multicolumn{6}{c|}{Fig. \ref{fig:cnameptrtestrecords}}
                                                   & \multicolumn{2}{c|}{Section \ref{fig:injectdot}}               & \multicolumn{2}{c|}{Section \ref{fig:inject000}} & Sec. \ref{fig:ldapopenjdk-crash} & Tab. \ref{tab:radsecproxy-exploits} & Fig. \ref{fig:libsfp2-payload}   \\%\multicolumn{3}{c|}{Usage share}
                                                   
    \cline{1-15} %
                                                   
    &\multirow{2}{*}{Test}                                           & \multirow{2}{*}{Base}   & \multirow{2}{*}{/}      & \multirow{2}{*}{@}      & \multirow{2}{*}{XSS}        & \multirow{2}{*}{SQL}        & \multirow{2}{*}{ANSI} 
                                                   & \multicolumn{2}{c|}{inject\textsubscript{\textbackslash.}} & \multicolumn{2}{c|}{inject\textsubscript{\textbackslash000}} & \multirow{2}{*}{LDAP} & \multirow{2}{*}{Eduroam} & \multirow{2}{*}{libspf2}  & \multirow{2}{*}{\cite{goingwild}} & \multirow{2}{*}{\cite{tatang2019large}} & \multirow{2}{*}{our study} \\ 
                                                   
    &                                               &        &                           &                          &            &            &
                                                    & CNAME & Direct       & CNAME & Direct &            &            & & & & \\ 
                                                    
\cline{1-15} %

\multirow{7}{*}{\rotatebox[origin=c]{90}{\scriptsize Recursive resolvers}}

&BIND (9.14.0)                                        & \cmark & \cmark & \cmark & \cmark & \cmark     & \cmark     & \xmark     & \xmark & \xmark     & \xmark & \cmark & \cmark & \cmark     & 60.2\%   & 70\%  & 11.4\% \\
&MaraDNS Deadwood (3.2.14)                            & \cmark & \cmark & \cmark & \cmark & \cmark     & \cmark     & \xmark     & \xmark & \xmark     & \xmark & \cmark & \cmark & \xmark$^2$ & -        & -     & - \\
&Unbound (1.9.1 )                                     & \cmark & \cmark & \cmark & \cmark & \cmark     & \cmark     & \xmark     & \xmark & \xmark     & \xmark & \cmark & \cmark & \cmark     & n/a      & 4.8\% & 1.9\% \\
&PowerDNS Recursor (4.3.0)                            & \cmark & \cmark & \cmark & \cmark & \cmark     & \cmark     & \xmark     & \xmark & \xmark     & \xmark & \cmark & \cmark & \cmark     & $>$3.2\% & 3\%   & 2.8\% \\
&Windows Server (2012 R2)                             & \cmark & \cmark & \cmark & \cmark & \cmark$^1$ & \cmark$^1$ & \xmark     & \xmark & \xmark     & \xmark & \cmark & \cmark & \cmark     & \multirow{3}{*}{ $ \Big\} >$2.5\% \hspace{6pt} } & \multirow{3}{*}{ 1\% } & \multirow{3}{*}{ $ \Big\} >$ 0.1\% \hspace{6pt} } \\
&Windows Server (2016)                                & \cmark & \cmark & \cmark & \cmark & \cmark     & \cmark     & \xmark     & \xmark & \xmark     & \xmark & \cmark & \cmark & \cmark     & & & \\
&Windows Server (2019)                                & \cmark & \cmark & \cmark & \cmark & \cmark     & \cmark     & \xmark     & \xmark & \xmark     & \xmark & \cmark & \cmark & \cmark     & & & \\

\cline{1-15}

\multirow{4}{*}{\rotatebox[origin=c]{90}{\scriptsize Forwarders}}

&pdnsd (1.2.9a)                                       & \cmark & \cmark & \cmark & \cmark & \cmark     & \cmark     & \xmark     & \xmark & \xmark     & \xmark & \cmark & \cmark & \cmark     & - & - & - \\
&dnsmasq (2.79)                                       & \cmark & \cmark & \cmark & \cmark & \cmark     & \cmark     & \xmark     & \xmark & \xmark     & \xmark & \cmark & \cmark & \cmark     & $>7.7$\% & 20\% & 42.5\%\\
&NxFilter (4.3.3.9)                                   & \cmark & \cmark & \cmark & \cmark & \cmark     & \cmark     & \xmark     & \xmark & \xmark     & \xmark & \cmark & \cmark & \cmark     & - & - & - \\
&systemd resolved (237)                               & \cmark & \cmark & \cmark & \cmark & \cmark     & \cmark     & \xmark     & \xmark & \xmark     & \xmark & \cmark & \cmark & \cmark     & - & - & - \\

\cline{1-15}

\multirow{11}{*}{\rotatebox[origin=c]{90}{\scriptsize Public resolvers}}

&OpenDNS                                              & \cmark & \cmark & \cmark & \cmark & \cmark$^1$ & \cmark$^1$ & \xmark     & \xmark & \xmark     & \xmark & \cmark & \cmark & \cmark      & - & - & 1.6\% \\
&Cloudflare Public DNS                                & \cmark & \cmark & \cmark & \cmark & \cmark$^1$ & \cmark$^1$ & \xmark     & \xmark & \xmark     & \xmark & \cmark & \cmark & \cmark      & - & - & 2.4\% \\
&Comodo Secure DNS                                    & \cmark & \cmark & \cmark & \cmark & \cmark     & \cmark     & \xmark     & \xmark & \xmark     & \xmark & \cmark & \cmark & \cmark      & - & - & - \\
&Google Public DNS                                    & \cmark & \cmark & \cmark & \cmark & \cmark     & \cmark     & \xmark     & \xmark & \xmark     & \xmark & \cmark & \cmark & \cmark      & - & - & 17.1\% \\
&Hurricane Electric                                   & \cmark & \cmark & \cmark & \cmark & \cmark     & \cmark     & \xmark     & \xmark & \xmark     & \xmark & \cmark & \cmark & \cmark      & - & - & - \\
&Neustar UltraRecursive                               & \cmark & \cmark & \cmark & \cmark & \cmark     & \cmark     & \xmark     & \xmark & \xmark     & \xmark & \cmark & \cmark & \cmark      & - & - & - \\
&Norton ConnectSafe                                   & \cmark & \cmark & \cmark & \cmark & \cmark     & \cmark     & \xmark     & \xmark & \xmark     & \xmark & \cmark & \cmark & \cmark      & - & - & - \\
&Oracle Dyn                                           & \cmark & \cmark & \cmark & \cmark & \cmark     & \cmark     & \xmark     & \xmark & \xmark     & \xmark & \cmark & \cmark & \cmark      & - & - & - \\
&SafeDNS                                              & \cmark & \cmark & \cmark & \cmark & \cmark     & \cmark     & \xmark     & \xmark & \xmark     & \xmark & \cmark & \cmark & \cmark      & - & - & - \\
&VeriSign Public DNS                                  & \cmark & \cmark & \cmark & \cmark & \cmark     & \cmark     &(\xmark)$^5$& \xmark & \xmark     & \xmark & \cmark & \cmark & \cmark      & - & - & - \\
&Yandex DNS                                           & \cmark & \cmark & \cmark & \cmark & \cmark     & \cmark     & \xmark     & \xmark & \xmark     & \xmark & \cmark & \cmark & \cmark      & - & - & - \\

\cline{1-15}

\multirow{10}{*}{\rotatebox[origin=c]{90}{\scriptsize Stub resolvers}}

& glibc    & \cmark     & \xmark       & \xmark       & \xmark       & \xmark       & \xmark       & \xmark     & - & \xmark     & - & \multicolumn{3}{c|}{ \multirow{10}{*}{ \makecell{ \Large n/a  \\  \scriptsize no SRV/TXT support \\ \scriptsize or does not apply } } }     \\ 
& musl     & \cmark     & \xmark       & \xmark       & \xmark       & \xmark       & \xmark       &(\xmark)$^5$& - & \xmark     & - &     &     &     \\ 
& dietlibc & \cmark     & \cmark       & \cmark       & \cmark       & \cmark       & \cmark       &(\xmark)$^5$& - &(\xmark)$^5$& - &     &     &     \\ 
& uclibc   & \cmark     & \cmark       & \cmark       & \cmark       & \cmark       & \cmark       &(\xmark)$^5$& - &(\xmark)$^5$& - &     &     &     \\ 

& Windows  & \cmark     & \cmark       & \cmark       & \cmark       & \cmark       & \cmark       &(\xmark)$^5$& - &(\xmark)$^5$& - &     &     &     \\ 
& NetBSD   & \cmark     & \cmark       & (\cmark)$^3$ & \cmark       & \cmark       & (\cmark)$^3$ & \xmark$^3$ & - & \xmark$^3$ & - &     &     &     \\ 
& Mac OS X & \cmark     & \cmark       & \cmark       & \cmark       & \cmark       & (\cmark)$^3$ & \xmark$^3$ & - & \xmark$^3$ & - &     &     &     \\ 

& go*      & \cmark     & \cmark       & \cmark       & \cmark       & \xmark       & \cmark       &(\xmark)$^5$& - & \xmark     & - &    &    &    \\ 
& nodejs   & \cmark     & \cmark       & \cmark       & \cmark       & \cmark       & \cmark       & \xmark$^3$ & - &(\xmark)$^5$& - &  &    &    \\
\cline{3-12}
\cline{13-13}

& openjdk8*& \multicolumn{10}{c|}{\scriptsize InetAddress.getCanonicalHostName() gives PTR instead of CNAME}                                     & \cmark & \multicolumn{2}{|c|}{} \\ 

\cline{1-15}
 
\multicolumn{2}{|l|}{
\multirow{2}{*}{ Open resolvers in Internet }
}
 & 100\%  & 96.6\% & 96.3\% & 96.4\% & 95.9\% & 96.3\% & 1.3\%  & 2.7\% & 3.6\% & 4.6\% & 99.6\% & 99.6\% & 56.1\% \\
\multicolumn{2}{|l|}{} & 1,328,146 & 1,283,447 & 1,279,573 & 1,279,835 & 1,273,710 & 1,279,100 & 45,596 & 51,049 & 82,589 & 88,864 & 1,322,529 & 1,322,203 & 745,599 \\

\cline{1-15}
        
    \end{tabular}
    \\ \cmark: Vulnerable. $^1$: record converted to lower case. $^2$: NXDOMAIN/no response. $^3$: output was escaped. $^5$: Record is misinterpreted, injection is not cached. \\ $*$:~Uses system stub resolver by default but offers a builtin one. \\
    \captionsetup{justification=centering}
    \caption{Forward-lookup test results for all groups of resolvers.}
    \label{tab:resolvertests}
    \vspace{-12pt}
\end{table*}

\begin{table}[ht]
    \footnotesize
    \centering
    \setlength\tabcolsep{4pt}
    \begin{tabular}{|r|c|cccc|ccc|}
    \hline
    
        {\bf Test}     & {\bf Base}       & /            &  @            & \textbackslash.     & \textbackslash000        & XSS        & SQL        & ANSI         \\ 
        \tiny Payload (Fig.\ref{fig:cnameptrtestrecords})       & \tiny 1.1.1.1    & \tiny 2.2.2.2      & \tiny 3.3.3.3      & \tiny 5.5.5.5             & \tiny 4.4.4.4                  & \tiny 6.6.6.6      & \tiny 7.7.7.7      & \tiny 8.8.8.8      \\ \hline
        glibc    & \cmark     & \xmark       & \xmark       & \xmark              & \xmark                   & \xmark     & \xmark     & \xmark       \\ \hline
        musl     & \cmark     & \cmark       & \cmark       & \cmark              & \cmark                   & \cmark     & \cmark     & \cmark       \\ \hline
        dietlibc & \cmark     & \cmark       & \cmark       & \cmark              & \cmark                   & \cmark     & \cmark     & \cmark       \\ \hline
        uclibc   & \cmark     & \cmark       & \cmark       & \cmark              & \cmark                   & \cmark     & \cmark     & \cmark       \\ \hline
        
        \hline
        
        windows  & \cmark     & \cmark       & \cmark       & \cmark              & \cmark                   & \cmark     & \cmark     & \cmark       \\ \hline
        netbsd   & \cmark     & \cmark       & (\cmark)$^2$ & (\cmark)$^2$        & (\cmark)$^2$             & \cmark     & \cmark     & (\cmark)$^2$ \\ \hline
        mac os x & \cmark     & \cmark       & \cmark       & (\cmark)$^2$        & \cmark                   & \cmark     & \cmark     & \cmark       \\ \hline

        \hline
        
        go*      & \cmark     & \cmark       & \cmark       & \cmark              & (\cmark)$^3$             & \cmark     & \cmark     & \cmark       \\ \hline
        openjdk8*   & \cmark     & \xmark       & \cmark       & (\cmark)$^2$        & (\cmark)$^3$             & \cmark$^4$ & \cmark   & \cmark         \\ \hline
        node     & \cmark     & \cmark       & \cmark       & (\cmark)$^2$        & \cmark                   & \cmark     & \cmark     & \cmark       \\ \hline

    \end{tabular}
    \\ \cmark:~Vulnerable. $^2$:~output was escaped. $^3$:~Zero-byte did not stop output. \\ $^4$:~Alternative XSS payload with \stt{" "} instead of \stt{"/"}. \\ $*$:~Uses system stub resolver by default but offers a builtin-one.
    \vspace{-5pt}
    \caption{Reverse-lookup results for different stub resolvers.}
    \label{tab:gethostbyaddr}
    \vspace{-12pt}
\end{table}

\section{Root Causes, Insights and Mitigations}\label{sc:mitigations}

\subsection*{Missing Specifications}
The study that we carried out in this work showed us that one of the root factors allowing our attacks is a lack of threat model in the standard RFCs as well as a lack of specifications on DNS and on its interactions with applications:

{\bf Threat modelling.} There is a lack of threat modelling in the DNS infrastructure and in the interaction of DNS with the applications. The RFCs should provide a threat model discussing potential pitfalls. %
For instance, most applications typically expect hostnames, but in the RFCs, this is not considered.

{\bf DNS record parsing.} There is a lack of detailed specification on how to parse DNS records. This critical functionality should be specified in the RFCs. %

{\bf Validation of DNS records.} There is a lack of  standardised implementation for validation of DNS records. This is important esp. for non-address record types as these are not supported by the OS resolver (e.g., libc). Similarly to, say libraries for generating DNSSEC keys, there should be an implementation for validation of received DNS records. %

{\bf Domain names vs hostnames.} There is a discrepancy between definitions of domain names and hostnames, which leads to confusion in DNS software in how to parse the DNS records. To avoid pitfalls the same rules should apply to both. %

\subsection*{Mitigations}

{\bf Applications.} Since DNS resolvers serve data from untrusted Internet sources, the applications should always treat data from DNS the same way they treat user input, hence validation of formatting \new{and escaping} should always be performed \new{(ie. validate html special characters, etc.)}, regardless if the API used to receive the data indicates an already checked result (POSIX) or not.

{\bf System stub resolvers.} Stub resolvers should be modified to check if domain names returned by POSIX calls like \stt{gethostbyname()} are valid hostnames, \cite{rfc1123}. If not, the domain name should not be given to the application\new{, like it is implemented in glibc already}. Merely escaping special characters as done by netbsd can still be vulnerable. For instance, we demonstrate this with the \stt{6.6.6.6.in-addr.arpa} Cross-Site Scripting payload in Figure~\ref{fig:cnameptrtestrecords} which does not contain any character typically escaped in domain names and, for example, can be used to execute an attack against OpenWRT. \new{Furthermore non-libc DNS libraries should follow the same rules (i.e., only allow hostnames as per \cite{rfc1123}) even for non-standard lookup types like SRV to prevent confusion among developers who only used libc resolvers before. Guidance on how to implement such checks should ideally be given by standardisation bodies.}

{\bf DNS resolvers.} Filtering DNS responses on the DNS resolver or forwarder level is possible but is against the DNS standard \cite{rfc3597,rfc1035,rfc1123}. Changing this requires a discussion in the corresponding working groups within the IETF, which we are initiated within our disclosure efforts. Nevertheless, performing checks on DNS records is challenging: some applications, like SRV service discovery \cite{rfc2782}, require domain names with characters that are not allowed in hostnames (e.g., underscore). Defining a list of allowed characters so that legitimate applications would still work but injection attacks would be blocked should be further investigated and is not straightforward. In particular, it is difficult to foresee what characters and formats will be needed by future applications, hence a `too-restrictive' list of allowed characters would make DNS less transparent, possibly introducing obstacles in deployment of new applications, or when adding new versions or new features to existing applications. On the other hand, leaving this completely transparent may lead to confusion about what values a field can actually have, and can even introduce vulnerabilities -- our work can be generalised to other Internet protocols. We show that the decision to enable easy future deployment of new applications by not restricting the domain names to alphanumeric characters exposes to attacks.

\new{Nevertheless, as an immediate protection against our attacks, operators which are unable to implement changes on the application- or stub-resolver-level might use filtering proxies which implement those validation steps on the network-level instead\footnote{We provide a proof-of-concept implementation of such a proxy at \url{https://xdi-attack.net}.}.} %

\new{{\bf Mitigations against cache-poisoning.} Since the cache-poisoning attacks cannot be reliably detected by downstream forwarders, these attacks must be mitigated by patching the resolvers causing the misinterpretation. If the resolver is only misinterpreting malicious records (but not caching them, like Verisign Public DNS), switching to a DNS forwarder which does not cache cross-CNAME records can prevent the attack. This however does not fix the root cause of the issue. When the resolver operator cannot not fix the vulnerability, switching to another DNS resolver is the best option.}

\section{Related Work}\label{sc:works}
 
 {\bf DNS cache poisoning.} Kaminsky provided the first demonstration of DNS cache poisoning attack \cite{Kaminsky08}. Since then DNS resolvers have been patched to support best practices [RFC5452] \cite{rfc5452}: randomising fields in requests, such as source port and DNS TXID, and validating them in responses, and also to apply checks such as bailiwick \cite{rfc2181}. This makes DNS resilient to off-path cache poisoning attacks. Nevertheless, recent works developed cache poisoning attacks when DNS responses are served over UDP \cite{gilad2013off}. The attacks use different side channels to predict the randomisation parameters, as well as other methodologies like fragmentation to bypass guessing the parameters altogether, \cite{herzberg2012security,cns:frag:dns,herzberg2013socket,herzberg2013vulnerable,brandt2018domain,stub:cache:infocom,zheng2020poison,qian:ccs20}. Our attacks are not limited by the transport protocol and apply to DNS over TCP as well as DNS over UDP. In contrast to all existing DNS cache poisoning attacks which evaluate the cache poisoning on one victim DNS resolver and then check if some selected population of DNS resolvers have the properties that could potentially make them vulnerable, our attack is the first to have been fully automated and evaluated on a large set of target networks, 3M, and the first to have been successfully launched against 105K resolvers. Prior attacks cannot be automated since they need to be tailored per each target victim resolver \cite{cns:frag:dns,brandt2018domain,stub:cache:infocom,zheng2020poison,qian:ccs20}. For instance, the servers set the fragmentation offset slightly differently hence making fragmentation difficult to match, the servers randomise the records in responses making the UDP checksum extremely difficult to match, UDP ports need to be measured per each target separately, overwriting cached records with new values depends on already cached records and caching policies, and so on. Our attack is not restricted by these hurdles. %
 
None of the proposed non-cryptographic defences prevent our cache poisoning attacks. Even the cryptographic protection with DNSSEC [RFC4033-RFC4035] \cite{rfc4033,rfc4034,rfc4035}, which blocks all previous DNS cache poisoning attacks, does not prevent our attacks in common settings. Furthermore, DNSSEC deployments were showed to often use weak cryptographic algorithms or vulnerable keys, \cite{dai2016dnssec,shulman2017one,chung2017understanding}. Cipher-suite negotiation schemes were proposed to allow easy adoption of stronger cryptographic ciphers \cite{herzberg2014less}.
 
Recent proposals for encryption of DNS traffic, such as DNS over HTTPS \cite{hoffman2018rfc} and DNS over TLS \cite{hu2016rfc}, although vulnerable to traffic analysis \cite{shulman2014pretty,siby2019encrypted}, may also enhance resilience to cache poisoning but do not prevent our injection attack. 
 
{\bf User input injections in web applications.} Injection vulnerabilities \cite{su2006essence} are the primary medium for performing remote exploits, including SQL injection attacks \cite{su2006essence}, Cross Site Scripting (XSS) \cite{nadji2009document}, buffer overflow \cite{dalton2008real}, XPath injections \cite{blasco2007introduction}, LDAP injections \cite{alonso_ldap_2008}, HTTP header injection \cite{johns2006requestrodeo}, Email header injection \cite{chandramouli2018measuring}, SMTP injection \cite{terada2015smtp}. All these differ from our \name\ attacks. User injection attacks via the web interfaces are typically blocked as user input is sanitised prior to being accepted by applications. Our attacks apply even when user input is properly validated and also where the users cannot provide any meaningful input at all, since we deliver malicious payloads by encoding them into DNS records. %

{\bf DNS rebinding attacks.} DNS rebinding attack \cite{jackson2009protecting,acar2018web} uses a script on the victim network and an external attacker to create a confusion in web browsers bypassing Same Origin Policy (SOP), say by mapping the external attacker to an internal IP address. This allows the attacker to impersonate internal hosts in order to bypass filtering that is applied on external packets, e.g., for spam or for Denial of Service (DoS) attacks. Our attacks target the internal services directly, without impersonation of internal devices. DNS rebinding are prevented, e.g., with filtering private IP addresses and blocking the resolution of external hostnames into internal IP addresses, or via DNS pinning \cite{jackson2009protecting} in web browsers - none of which prevent our attacks.

\section{Conclusions}\label{sc:conclusions}
Our work shows that central transparency-related principles of development of Internet systems should be reconsidered: 

{\bf Flexibility.} Be strict when sending and permissive when receiving is a good principle in the Internet. Systems and protocols that are too rigid are much more difficult to use and require significant changes to the existing infrastructure for adoption of new technologies or mechanisms. The huge success of DNS in providing platform to new applications is thanks to its transparent handling of DNS records. If DNS is made less transparent, e.g., by requiring that the records are checked for invalid characters, it would make the roll out of new applications in the Internet much more challenging. For instance, if DNS parsed each record, new applications using records containing not yet supported characters, e.g., not just alphanumeric characters, like in SRV record type, would require changes to the DNS servers all over the Internet to enable support for new characters. Unupgraded servers would risk failures or even crashes when processing the new records. On the other hand, making systems too tolerant can expose to vulnerabilities. We showed that leaving the specification completely open exposes DNS and the applications using DNS to attacks. Hence, a balance should be found between the ease of deployment and security.

{\bf Layering.} Although it is a known networking principle that each layer provides services to the layer above it, and the upper layer does not have to worry about the data provided by lower protocols, we show that when it comes to security this principle may result in vulnerabilities. We recommend that the validation of DNS data is integrated into applications directly not relying on the lower layers to do this for them. For instance, it may not always be possible for DNS to predict all the applications of the data that it provides and scenarios where it will be used. Hence even if DNS is changed to apply checks over the data in DNS records, the applications should nevertheless do the validation also themselves.

\section*{Acknowledgements}
We are grateful to Yuval Yarom and to the anonymous referees for their thoughtful feedback on our work.

This work has been co-funded by the German Federal Ministry of Education and Research and the Hessen State Ministry for Higher Education, Research and Arts within their joint support of the National Research Center for Applied Cybersecurity ATHENE and by
the Deutsche Forschungsgemeinschaft (DFG, German Research Foundation) SFB~1119. %

{
\footnotesize
\bibliographystyle{IEEEtran}
\bibliography{main.bib,rfc2.bib}
~
}

\end{document}